\documentclass[12pt,a4paper]{article}

\usepackage[utf8]{inputenc}
\usepackage[T1]{fontenc}
\usepackage[english]{babel}
\usepackage{authblk}
\usepackage[a4paper,margin=1in]{geometry}

\usepackage{newtxtext, newtxmath} 
\usepackage{setspace} 
\usepackage{microtype}

\usepackage{graphicx}

\usepackage{amssymb}
\usepackage{booktabs}

\usepackage{natbib}

\usepackage{amsmath,amssymb}
\usepackage{algorithm}
\usepackage{algpseudocode} 
\usepackage{enumitem}
\usepackage{amsmath,amssymb}
\usepackage{nccmath}
\usepackage{float}  

\usepackage{multirow}         
\usepackage{threeparttable}   

\usepackage{titlesec}
\titleformat{\subsubsection}
  {\normalfont\bfseries}
  {\thesubsubsection}
  {1em}
  {\boldmath}
\usepackage{placeins}

\usepackage{natbib}

\usepackage[colorlinks=false]{hyperref}

\title{Exchangeable Gaussian Processes for Staggered-Adoption Policy Evaluation}

\author[1]{Hayk Gevorgyan}
\author[2]{Konstantinos Kalogeropoulos}
\author[1]{Angelos Alexopoulos}

\affil[1]{Department of Economics, Athens University of Economics and Business, Athens, Greece}
\affil[2]{Department of Statistics, The London School of Economics and Political Science, London, United Kingdom}

\date{\today}

\begin{document}

\maketitle
\newcounter{chapter}
\renewcommand\thechapter{\arabic{chapter}}

\makeatletter
\@addtoreset{section}{chapter}
\makeatother
\renewcommand\thesection{\thechapter.\arabic{section}}

\newcommand{\chapter}[1]{%
  \par              
  \vspace{2em}      
  \refstepcounter{chapter}%
  \noindent\textbf{\Large \thechapter\quad #1}%
  \par\medskip      
}
\makeatletter
\renewcommand\section{\@startsection{section}{1}{\z@}%
  {-3.5ex \@plus -1ex \@minus -.2ex}%
  {2.3ex \@plus .2ex}%
  {\normalfont\large\bfseries}}
\makeatother

\begin{abstract}
We study the use of exchangeable multi-task Gaussian processes (GPs) for causal inference in panel data, applying the framework to two settings: one with a single treated unit subject to a once-and-for-all treatment and another with multiple treated units and staggered treatment adoption. Our approach models the joint evolution of outcomes for treated and control units through a GP prior that ensures exchangeability across units while allowing for flexible nonlinear trends over time. The resulting posterior predictive distribution for the untreated potential outcomes of the treated unit provides a counterfactual path, from which we derive pointwise and cumulative treatment effects, along with credible intervals to quantify uncertainty.

We implement several variations of the exchangeable GP model using different kernel functions. To assess prediction accuracy, we conduct a placebo-style validation within the pre-intervention window by selecting a ``fake'' intervention date. Ultimately, this study illustrates how exchangeable GPs serve as a flexible tool for policy evaluation in panel data settings and proposes a novel approach to staggered-adoption designs with a large number of treated and control units.

\end{abstract}

\chapter{Introduction}
\label{ch:introduction}

Modern public policy evaluation increasingly relies on longitudinal datasets that follow multiple units over time, such as states, regions, firms or individuals. In many applications, the central question is to assess the causal effect of a policy intervention that affects a subset of units at a known time, using the remaining units as controls to reconstruct the counterfactual trajectory that would have been observed, if an intervention never occurred. 

A prominent framework for such problems is the synthetic control approach \citep{abadie2003,AbadieDiamondHainmueller2010}, where the counterfactual path of the treated unit is estimated by a weighted sum of control units. This and related methods, including factor-model-based generalized synthetic control \citep{xu2017gsc} and Bayesian structural time-series models \citep{brodersen2015}, have been widely applied in economics and public policy. These approaches typically rely on assumptions about the specific functional form of the evolution of outcomes over time and on linear combinations of control units, which may be restrictive when trends are nonlinear, covariate effects are complex, or the dynamics differ substantially across units.

Gaussian process (GP) priors \citep{rasmussen2006} offer a flexible, non-parametric or semi-parametric alternative for modeling time-series dynamics and covariate relationships. They allow us to characterize relationships across different units, while remaining unrestricted in functional forms. Recent work has introduced GP-based formulations for counterfactual prediction and causal effect estimation in panel data settings \citep{giudice2022, benmichael2022multitaskgp}, demonstrating that GPs can provide a reliable way to quantify uncertainty about the counterfactual path, and they can accommodate nonlinear trends and time-varying relationships. Within the broad class of GPs, \emph{exchangeable Gaussian processes} have recently been proposed as a multi-task GP framework \citep{leroy2022magma, bouranis2023} , where a GP prior is placed on the latent mean process shared across units, and each unit-specific deviation is itself a GP, creating an exchangeable structure across units.

This paper adapts and extends the exchangeable GP framework to the context of causal inference with panel data and illustrates its use in two applications. The first is the Proposition~99 (Prop99) setting, where units correspond to U.S. states observed annually and the treated unit is California; the intervention associated with Prop99 is introduced for California after 1988, while the remaining states serve as controls. The second dataset is a large panel of Greek petrol stations observed at weekly frequency over the course of a year. A policy intervention affects a subset of stations at different points in time, while the remaining stations serve as never-treated controls. In both settings, our goal is to assess whether the intervention had a causal impact on the outcome by constructing counterfactual trajectories for treated units using the outcomes of non-treated units and covariates as predictors.

The paper is organized as follows. Section~\ref{ch:causal-setup} introduces the causal inference framework, the corresponding assumptions, and a review of state-of-the-art methods. Section~\ref{ch:exchangeable-gp} describes the exchangeable GP model. Section~\ref{ch:staggered-adoption} details the model-fitting and inference protocol for staggered adoption setting. Section~\ref{ch:estimation} presents the estimation details, including hyperparameter learning and the construction of posterior predictive counterfactual paths. Section~\ref{ch:application} presents the datasets, specifies the empirical design, and reports the results of the validation and policy-effect analyses.

\chapter{Causal Inference with panel data}
\label{ch:causal-setup}

In this section we formalize the causal inference problem in a panel time-series setting and introduce the estimands that will be reported in the final analysis. We present the potential outcomes framework specific to panel data with a policy intervention that affects a single units at a specific point in time.

\section{Panel structure and potential outcomes}
\label{sec:panel-notation}

Consider a panel with units (e.g., states, firms) indexed by $i = 1,\dots,m$ and discrete time periods indexed by $t = 1,\dots,T$. Let $\mathcal{I} = \{1,\dots,m\}$ denote the set of units and $\mathcal{T} = \{1,\dots,T\}$ the set of time periods. For each unit and time, we observe an outcome $y_{it} \in \mathbb{R}$ of interest and possibly a
vector of observed covariates $x_{it} \in \mathbb{R}^p$.

A policy intervention is introduced at different unit-specific time points, and it affects a subset of units $\mathcal{I}_1 \subseteq \mathcal{I}$. The remaining units $\mathcal{I}_0 = \mathcal{I}\setminus \mathcal{I}_1$ serve as potential controls. In the remainder of this section, we focus on the case of \emph{a single treated unit} with a sharp intervention at time $T_0$, which corresponds to the setting of our empirical application with a single treated unit (California) and is also a simplification for our applied case of Greek businesses with staggered treatment adoption.

Let $i^\star$ denote the treated unit, so that $\mathcal{I}_1 = \{i^\star\}$ and $\mathcal{I}_0$ contains all other units. We partition the time axis into a pre-intervention period
$\mathcal{T}_0 = \{1,\dots,T_0\}$ and a post-intervention period
$\mathcal{T}_1 = \{T_0+1,\dots,T\}$.

For each unit $i$ and time $t$, define a treatment indicator
$w_{it} \in \{0,1\}$, where $w_{it} = 1$ if the policy has already been applied to unit $i$ at time $t$, and $w_{it} = 0$ otherwise. In the case of a single treated unit with a single strict intervention time we have $w_{it} = 0$ for all $i \neq i^\star$ and $t$, and
\[
w_{i^\star t} =
\begin{cases}
0, & t \leq T_0,\\
1, & t > T_0.
\end{cases}
\]

Within the potential outcomes framework, for each $(i,t)$ there are two potential outcomes: $y_{it}(0)$ denotes the outcome that would be observed for unit $i$ at time $t$ in the absence of treatment, and $y_{it}(1)$ denotes the outcome that would be observed under
treatment. The actual outcome is
\[
y_{it}
= y_{it}(0)\,\mathbb{I}\{w_{it} = 0\}
+ y_{it}(1)\,\mathbb{I}\{w_{it} = 1\},
\]
where $\mathbb{I}\{\cdot\}$ is an indicator function. The fundamental problem is that for the treated unit $i^\star$, we only observe $y_{i^\star t}(1)$, and not $y_{i^\star t}(0)$ for any given $t > T_0$. Hence, our main task is to estimate the sequence $\{y_{i^\star t}(0): t \in \mathcal{T}_1\}$, which yields the
counterfactual path for the treated unit.

\section{Causal estimands}
\label{sec:estimands}

Our primary objective is to quantify the causal effect of the policy on the outcome of the treated unit over time, together with the uncertainty around these effects.

For each unit $i$, the individual causal effect at time $t$ is denoted as
\[
\delta_{i,t} = y_{i,t}(1) - y_{i,t}(0).
\]

In our empirical applications, we consider both the Proposition~99 (Prop99) setting with a single treated unit $i^\star$ (California) and a Greek business-level panel with multiple treated units. In the Prop99 application, we aim to estimate $\delta_{i^\star,t}$ for all post-intervention times $t \in \mathcal{T}_1 = \{T_0+1,\dots,T\}$. In the Greek business application, we are interested in estimating $\delta_{i,t}$ for all treated units $i \in \mathcal{I}_1$ at all post-intervention times.

Denote the set of outcomes and covariates that are not affected by the treatment as $\mathcal{D}_0$. This includes the covariates of all units, including the treated unit, and the set of outcomes of untreated units, plus the set of pre-treatment outcomes of the treated unit. Given $\mathcal{D}_0$, we define the mean of the individual causal effect at time $t$
as
\begin{equation}
\tau_{i^\star,t}
= \mathbb{E}\bigl[\delta_{i^\star,t} \mid \mathcal{D}_0\bigr],
\qquad t \in \mathcal{T}_1.
\label{eq:additive-effect}
\end{equation}

We are also interested in the uncertainty surrounding the causal effect. This can be quantified via the posterior variance
\begin{equation}
\sigma_{i^\star,t}^2
= \mathbb{V}\bigl[\delta_{i^\star,t} \mid \mathcal{D}_0\bigr],
\label{eq:var-effect}
\end{equation}
or, equivalently, through posterior quantiles
\begin{equation}
q_{\delta_{i^\star,t}}(\alpha)
= F_{\delta_{i^\star,t}}^{-1}\bigl(\alpha \mid \mathcal{D}_0\bigr),
\qquad \alpha \in (0,1),
\label{eq:quantiles-effect}
\end{equation}
where $F_{\delta_{i^\star,t}}$ is the posterior cumulative distribution function of $\delta_{i^\star,t}$. We generally report central $95\%$ credible intervals together with the posterior mean
$\tau_{i^\star,t}$ for each $t$.

Within a Gaussian process framework with Gaussian likelihood, these quantities are straightforward to derive. The posterior predictive distribution of the counterfactual outcome $y_{i^\star,t}(0)$ given
$\mathcal{D}_0$ and $x_{i^\star,t}$ is Gaussian, with some mean
$m_{0,t}$ and variance $v_{0,t}^2$. The observed treated outcome $y_{i^\star,t}(1)$ is fixed, so the difference $ \delta_{i^\star,t} = y_{i^\star,t}(1) - y_{i^\star,t}(0) $
is also Gaussian under the posterior, with
\[
\delta_{i^\star,t} \mid \mathcal{D}_0
\sim \mathcal{N}\bigl(\tau_{i^\star,t}, \sigma_{i^\star,t}^2\bigr),
\]
where $\tau_{i^\star,t} = y_{i^\star,t}(1) - m_{0,t}$ and $\sigma_{i^\star,t}^2 = v_{0,t}^2$. Thus point estimates and credible intervals for the time-specific effects are derived directly.

In addition to the impacts specific to each time point, we are also interested in the aggregate effects of the treatment. For the treated unit $i^\star$ we define the cumulative treatment effect as
\begin{equation}
\Delta_{i^\star}
= \sum_{t = T_0+1}^{T} \delta_{i^\star,t},
\label{eq:cumulative-effect}
\end{equation}
which represents the total change in outcome variable caused by the treatment. Another commonly reported summary is the average treatment effect of the treated unit,
\begin{equation}
\delta_{i^\star}
= \frac{1}{T - T_0} \sum_{t = T_0+1}^{T} \delta_{i^\star,t}
\label{eq:average-effect}
\end{equation}
which measures the average impact of the intervention for the post-treatment period.

We define the corresponding expected average and cumulative treatment effects as
$\tau_{i^\star} = \mathbb{E}\!\left[\delta_{i^\star}\mid \mathcal{D}_0\right],
\label{eq:expected-cumulative-effect}$ and $
C_{i^\star} = \mathbb{E}\!\left[\Delta_{i^\star}\mid  \mathcal{D}_0\right],
\label{eq:expected-average-effect}
$
respectively.

In the empirical analysis, we report time-specific summaries $\tau_{i,t}$ together with cumulative and average effects and their associated posterior uncertainty intervals. For Prop99 these summaries refer only to California, while for the Greek business application they are also aggregated across treated firms.

\section{Assumptions}
\label{sec:assumptions}

The estimands defined in Section~\ref{sec:estimands} require assumptions to guarantee that the differences in the potential outcome paths are a direct consequence of the policy intervention. In this section we state the main assumptions for the analysis. Some of these assumptions cannot be formally tested and must instead be justified for a given application. We adapt the set of assumptions, as used in \citep{manchettietal}.

\paragraph{Assumption 1 (Single sharp intervention).}
There exists a time $T_{0} \in \{1,\dots,T\}$ such that, for the treated unit $i^\star$, the treatment indicator satisfies $w_{i^\star t} = 0$ for all $t \leq T_{0}$ and $w_{i^\star t} = 1$ for all $t > T_{0}$. For all control units $i \in \mathcal{I}_0$, $w_{it} = 0$ at all times $t$. This assumption states that the treated unit experiences a single intervention and that the treated status remains unchanged for the whole post-intervention period. This allows us to work in a framework where there is a clear distinction between treated and non-treated units, as well as the corresponding time intervals.

\paragraph{Assumption 2 (Temporal no-interference).} For each unit $i$ and time $t$, let $y_{it}(w_{i,1:t})$ denote the potential outcome when the treatment history of unit $i$ up to time $t$ is $w_{i,1:t}$, and let $y_{it}(w_{1:m,1:t})$ denote the potential outcome when the full treatment history of all units up to time $t$ is $w_{1:m,1:t}$. We assume that
\[
y_{it}(w_{i,1:t})
= y_{it}(w_{1:m,1:t})
\quad\text{for all } i,t \text{ and all treatment paths } w_{1:m,1:t}.
\]
In other words, the outcome of unit $i$ at time $t$ depends only on its own treatment path up to time $t$, and not on the treatment history of other units.

\paragraph{Assumption 3 (Covariates unaffected by treatment).}
Let $x_{it}(0)$ and $x_{it}(1)$ denote the potential covariate values that would be observed for unit $i$ at time $t$ when untreated and treated, respectively. For all units and all $t \geq T_{0}$,
\[
x_{it}(1) = x_{it}(0).
\]
This assumption states that the covariates used in the model are not affected by the treatment. These variables are chosen specifically to improve prediction of the counterfactual path. If they were influenced by the treatment, conditioning on them could introduce bias.

\paragraph{Assumption 4 (Non-anticipation of outcomes).}
We assume that for all $t < T_{0}$ and for any two possible future paths $w_{i,t+1:T}$ and
$w'_{i,t+1:T}$,
\[
y_{it}(w_{i,1:t}, w_{i,t+1:T})
= y_{it}(w_{i,1:t}, w'_{i,t+1:T}).
\]
This implies that before the intervention time $T_{0}$ the outcome of unit $i$ at time $t$ depends only on its past and current treatment history $w_{i,1:t}$ and is unaffected by how the treatment will be assigned to that unit in the future. More precisely, the assumption states that the policy intervention that will happen in the future cannot affect the outcome variable now.

\paragraph{Assumption 5 (Non-anticipating treatment).}
The treatment assignment at time $T_{0}$ for unit $i$ depends only on its past covariates and past outcomes,
\[
p\bigl(w_{iT_{0}} \mid w_{1:m,1:T_{0}-1}, y_{1:m,1:T}, x_{1:m,1:T}\bigr)
= p\bigl(w_{iT_{0}} \mid y_{i,1:T_{0}-1}, x_{i,1:T_{0}-1}\bigr).
\]
This assumption indicates that there are no additional unobserved factors that jointly affect the decision to treat at time $T_{0}$, other than the past covariates and outcome values of the unit.

Taken together, these assumptions allow us to work with the simplified notation $y_{it}(0)$ and $y_{it}(1)$, introduced in Section~\ref{sec:panel-notation}, without referring to the use of treatment histories. This means that all the observations that are unaffected by the treatment can be used to learn about the distribution of the untreated potential outcomes for the treated unit in the post-intervention period.

\section{State-of-the-art methods for panel data causal inference}
\label{sec:classical-estimators}

In panel data policy evaluation settings, we aim to infer the distribution of the untreated potential outcomes $\{y_{i^\star t}(0): t \in \mathcal{T}_1\}$ for the treated unit, given only its pre-intervention outcome trajectory, the full trajectories of the control units, and the assumptions in Section~\ref{sec:assumptions}. Classical approaches to causal inference can be seen as different ways of specifying a model for the joint evolution of $\{y_{it}(0)\}$ across units and time, under which the counterfactual path of the treated unit becomes identifiable.

A sensible starting point is the difference-in-differences (DiD) framework in its linear two-way fixed effects (TWFE) form \citep[see, e.g.,][]{angristpischke2009}
\[
y_{it} = \alpha_i + \gamma_t + \beta w_{it} + \varepsilon_{it},
\]
where $\alpha_i$ are unit-specific fixed effects, $\gamma_t$ are time-specific fixed effects, $w_{it}$ is the treatment indicator, and $\beta$ captures the effect of the treatment. This modeling structure imposes an additive assumption on the outcome variable with no interaction between unit and time-specific components. The parameter $\beta$ can be interpreted as an average treatment effect. When treatment effects change over time, or when trends are nonlinear, this method can be too restrictive and can lead to biased or hard-to-interpret estimates.

Another widely used method, the synthetic control \citep{abadie2003,AbadieDiamondHainmueller2010}, is specified for a single treated unit with several controls and a relatively long pre-intervention period. In this modeling approach, a set of non-negative weights on the control units are estimated that minimize the discrepancy between the treated unit and a weighted average of controls in the pre-intervention period. If such a weighted combination reproduces well the pre-intervention trajectory of the treated unit, the same combination is assumed to approximate the untreated potential outcomes in the post-intervention period. A related approach is synthetic difference-in-differences (SDID) \citep{arkhangelsky2021sdid}, which combines synthetic control-style weighting with a difference-in-differences adjustment.

Generalised synthetic control (GSC) \citep{xu2017gsc} extends the TWFE framework by allowing for interactions between time-specific and unit-specific effects,
\[
y_{it}(0) = \gamma_t + \lambda_t^\top \mu_i + \varepsilon_{it},
\]
where $\gamma_t$ is a common time effect, $\lambda_t$ is a vector of
unobserved factors, shared across units at each $t$, and $\mu_i$ is a vector of unit-specific loadings. The $\lambda_t^\top \mu_i$ term represents the interactive part, where time specific and unit-specific components interact. The factors and loadings are estimated from the control units and the pre-intervention data. The estimated loadings and factors are then used to obtain the treated unit's counterfactual trajectory.

Bayesian structural time-series (BSTS) models \citep{brodersen2015} offer a different perspective. In this case the outcomes of the treated unit are modeled through a state-space representation. In its extended form, it decomposes the untreated outcome into latent components, such as a local level, a trend, seasonal fluctuations and a regression term on the outcomes of control units. The latent states follow state transition equations, and priors are placed on the state variances and regression coefficients. Given the pre-intervention data, one can sample from the posterior of the latent states and parameters and propagate the model forward in time to obtain the predictive distribution of $y_{i^\star t}(0)$ for $t \in \mathcal{T}_1$.

The methods presented above share a common logic. Each imposes a structural form for the evolution of untreated outcomes across units and time, uses the pre-intervention (and optionally control) outcome data to learn the parameters or latent structure, and then uses the learnt coefficients to impute the counterfactual path for the treated unit. The reliability of the resulting conclusions depends on whether the imposed structure is sufficiently flexible to capture the outcome dynamics and the dependence between units.

\section{Gaussian Processes for Causal Inference in Panel Data}
\label{sec:motivation-gp}

Gaussian processes (GPs) provide a flexible Bayesian framework for constructing counterfactual outcomes in panel settings. Rather than imposing a parametric structure, GP-based approaches learn potentially nonlinear dynamics directly from the data, while producing predictive distributions that quantify uncertainty. 

Multi-task GP models treat each unit’s outcome trajectory as one ``task'' and learn them jointly, so information can be shared across units when forecasting the untreated path of treated units \citep{benmichael2022multitaskgp, giudice2022}. Practically, the model is trained on pre-intervention outcomes for treated units and on the full outcomes for never-treated controls, alongside corresponding potential covariates. Post-intervention counterfactual are then predicted through posterior predictive distribution. This joint modeling step plays a similar role to the ``borrowing strength'' of synthetic control methods, but does it in a nonparametric way and yields uncertainty intervals around counterfactuals.

A central modeling choice is the covariance kernel, that encodes which trajectories are expected to look similar and over what time scales. Kernels can be designed to capture smooth long-run movements, shorter-run fluctuations, and correlations across units, and can incorporate observed covariates to improve forecasting when covariate dynamics are informative for outcomes. In multi-task settings, kernels also determine how how strongly one unit can inform predictions for another, making kernel choice a key component for adapting GP counterfactual methods to different panel structures.

\chapter{Exchangeable Gaussian Processes for Panel Data Causal Inference}
\label{ch:exchangeable-gp}

In this section we present the Exchangeable Gaussian process methodology for panel data causal inference. We specify the model in a single treated unit setting, but it is directly extendable for the multiple treated unit case. The exchangeable GP model specifies a GP prior on a latent mean trajectory shared by all units and on unit-specific deviations around this mean. This method establishes an exchangeable covariance structure between units. Conditioned on the observed pre-intervention data, the posterior predictive distribution under the no-intervention scenario provides a counterfactual path for the treated unit, from which we can derive the causal estimands defined in Section~\ref{sec:estimands} together with uncertainty quantification.

Let each unit $i$ be observed at times $t = 1,\dots,T$, with scalar outcomes
\begin{equation}
\mathbf{y}_i(0) = \bigl(y_{i1}(0),\dots,y_{iT}(0)\bigr)^\top \in \mathbb{R}^T,
\end{equation}
where $y_{it}(0)$ denotes the \emph{untreated} (no-intervention) potential outcome at time $t$, and associated covariates
\begin{equation}
\mathbf{x}_i = \bigl(x_{i1},\dots,x_{iT}\bigr)^\top \in \mathbb{R}^{T \times p}.
\end{equation}
In the first part of this section we ignore covariates and focus on time as the only input. We will present the extension with covariates later in the section. For simplicity, we first assume a common time length across all units $i = 1,\dots,m$,
$t \in \{1,\dots,T\}$. We get the following stacked vector that combines untreated potential outcomes of all units.
\[
\mathbf{y}
=
\bigl(\mathbf{y}_1(0)^\top,\dots,\mathbf{y}_m(0)^\top\bigr)^\top \in \mathbb{R}^{mT}.
\]
The untreated potential outcome observation model is specified by the following equation
\begin{equation}
y_{it}(0) = f_i(t) + \varepsilon_{it},
\qquad
\varepsilon_{it} \sim \mathcal{N}(0,\omega_i^2),
\label{eq:exch-observation}
\end{equation}
with independent Gaussian noise across $(i,t)$ and $\varepsilon_{it} \;\perp\; f_j$ for all $i, j$. The treatment only affects the realized post-intervention observations for the treated unit, $i^\star$.

\paragraph{Decomposition into common and idiosyncratic components.}
The key idea of the exchangeable GP is to decompose each unit-specific latent function $f_i(t)$ into a common component shared by all units and an idiosyncratic component:
\begin{equation}
f_i(t) = \mu(t) + g_i(t),
\qquad i=1,\dots,m,\; t \in \mathcal{T}.
\label{eq:exch-decomposition}
\end{equation}
This is equivalent to a linear model of coregionalization \citep{alvarez2012kernels} with two latent processes and fixed loadings. All units load with weight $1$ on the common process $\mu$, and only unit $i$ loads
on its own deviation $g_i$, again with weight $1$.

We place independent GP priors on $\mu$ and the $g_i$:
\begin{align}
\mu(\cdot) &\sim \mathcal{GP}\bigl(0,\;\sigma_\mu^2\,k(\cdot,\cdot)\bigr),
\label{eq:prior-mu}\\[0.3em]
g_i(\cdot) &\stackrel{\text{i.i.d.}}{\sim} \mathcal{GP}\bigl(0,\;\sigma_g^2\,k(\cdot,\cdot)\bigr),
\qquad i=1,\dots,m.
\label{eq:prior-gi}
\end{align}
We assume
\[
\mu \;\perp\; g_i \;\perp\; g_j \quad (i\neq j), \qquad
(\mu,g_i) \;\perp\; \varepsilon_{i,t}.
\]
The kernel $k$ encodes temporal dependence. In our applications we will consider a \emph{Ornstein--Uhlenbeck (OU) kernel},
$k^{\mathrm{OU}}(t,s) = \exp\!\left(-\frac{|t-s|}{\ell}\right)$, and an \emph{RBF kernel}, 
$k^{\text{RBF}}(t,s) = \exp\!\left(-\frac{(t-s)^2}{2\ell^2}\right)$. Both kernels are assumed to have unit variance and lengthscale $\ell$,

The variance parameters $\sigma_\mu^2$ and $\sigma_g^2$ represent the contribution of the common component and the unit-specific deviations, respectively, while $\ell$ controls smoothness.

For each unit, define
$\boldsymbol{\mu} = \bigl(\mu(1),\dots,\mu(T)\bigr)^\top,
\;
\mathbf{g}_i = \bigl(g_i(1),\dots,g_i(T)\bigr)^\top,$
and denote 
\[
\mathbf{f}_i = \boldsymbol{\mu} + \mathbf{g}_i.
\]

Let $
\mathbf{f}
=
\bigl(\mathbf{f}_1^\top,\dots,\mathbf{f}_m^\top\bigr)^\top
\in \mathbb{R}^{mT}$, and let $K \in \mathbb{R}^{T\times T}$ be the covariance matrix with entries
$(K)_{ts} = k(t,s)$. 

From \eqref{eq:prior-mu}-\eqref{eq:prior-gi} we obtain $
\boldsymbol{\mu} \sim \mathcal{N}\bigl(0,\;\sigma_\mu^2 K\bigr),
\;
\mathbf{g}_i \stackrel{\text{i.i.d.}}{\sim} \mathcal{N}\bigl(0,\;\sigma_g^2 K\bigr), $
and by independence,
\[
\mathrm{Cov}\bigl(\mathbf{f}_i\bigr)
=
\bigl(\sigma_\mu^2 + \sigma_g^2\bigr) K.
\]

For $i \neq j$, we get $
\mathrm{Cov}\bigl(f_i(t),f_j(s)\bigr)
=
\sigma_\mu^2 k(t,s)$,
while for $i=j$, $
\mathrm{Cov}\bigl(f_i(t),f_i(s)\bigr) = \bigl(\sigma_\mu^2 + \sigma_g^2\bigr) k(t,s).
$
Hence the covariance of the stacked vector $\mathbf{f}$ has an $m \times m$ block structure
\begin{equation}
\mathrm{Cov}(\mathbf{f})
=
\begin{bmatrix}
(\sigma_\mu^2+\sigma_g^2) K & \sigma_\mu^2 K            & \cdots & \sigma_\mu^2 K \\
\sigma_\mu^2 K             & (\sigma_\mu^2+\sigma_g^2) K & \cdots & \sigma_\mu^2 K \\
\vdots                                   & \vdots                                   & \ddots & \vdots                         \\
\sigma_\mu^2 K             & \sigma_\mu^2 K             & \cdots & (\sigma_\mu^2+\sigma_g^2) K
\end{bmatrix},
\label{eq:exch-Kf-block}
\end{equation}
or, equivalently, in Kronecker form
\begin{equation}
\mathrm{Cov}(\mathbf{f})
=
(\mathbf{1}_m \mathbf{1}_m^\top) \otimes \bigl(\sigma_\mu^2 K\bigr)
\;+\;
I_m \otimes \bigl(\sigma_g^2 K\bigr),
\label{eq:exch-Kf-kronecker}
\end{equation}
where $\mathbf{1}_m$ is the $m$-vector of ones. The off-diagonal blocks are all identical and depend only on the common kernel, while the diagonal blocks contain the sum of common and unit-specific kernels. This shows the \emph{exchangeable} structure of the block covariance matrix: the joint distribution of $\mathbf{f}$ is invariant under permutations of the unit index
$i$.

Deriving the marginal likelihood is straightforward in the case of Gaussian noise. From \eqref{eq:exch-observation} we have 

\begin{equation}
    \mathbf{y}_i(0) = \mathbf{f}_i + \boldsymbol{\varepsilon}_i,
\qquad
\mathbf{y} = \mathbf{f} + \boldsymbol{\varepsilon},
\label{eq:observation-full-matrix}
\end{equation}
where
$\boldsymbol{\varepsilon} = (\boldsymbol{\varepsilon}_1^\top,\dots,\boldsymbol{\varepsilon}_m^\top) \sim \mathcal{N}\bigl(0,\;\Omega\bigr),\;$ $
\Omega = \mathrm{diag}\bigl(\omega_1^2 I_T,\dots,\omega_m^2 I_T\bigr).$

As $\varepsilon_{i,t}$ are mutually independent, and independent of $\mathbf{f}$, marginalizing over $\mathbf{f}$ yields
\begin{equation}
\mathbf{y} \sim \mathcal{N}\bigl(0,\;K_f + \Omega\bigr),
\qquad
K_f = \mathrm{Cov}(\mathbf{f})
\text{ given in \eqref{eq:exch-Kf-block}--\eqref{eq:exch-Kf-kronecker}}.
\label{eq:exch-marginal-y}
\end{equation}

In practice, we can learn the hyperparameters of the exchangeable GP by maximizing the log marginal likelihood (Type 2 ML, \citet{rasmussen2006}). The hyperparameters are
$(\sigma_\mu^2,\sigma_g^2,\ell,\{\omega_i^2\})$.

\paragraph{Extension with covariates.}
We now extend the model to allow the unit-specific deviations to depend on covariates, as well as on time. By extending the decomposition in \eqref{eq:exch-decomposition}, we get
\begin{equation}
f_i(t)
=
\mu(t) + g_{i,1}(t) + g_{i,2}(x_{it}),
\label{eq:exch-decomposition-cov}
\end{equation}
where $g_{i,1}$ is a time-only deviation, $g_{i,2}$ captures additional unit-specific variation explained by covariates, and $x_{it} \in \mathbb{R}^p$
denotes the covariates of unit $i$ at time $t$. An extension of the model that includes covariates as shared input follows directly by adding another $\mu(x_t)$ function with global covariates $x_t$. This specification is not described in this section, but it will be applied in the Prop99 setting in Section~\ref{sec:prop99}. 

We place independent GP priors:
\begin{align}
\mu(\cdot) &\sim \mathcal{GP}\bigl(0,\;\sigma_\mu^2\,k_{\text{time}}(\cdot,\cdot)\bigr),
\\[0.2em]
g_{i,1}(\cdot) &\stackrel{\text{i.i.d.}}{\sim} \mathcal{GP}\bigl(0,\;\sigma_{g_1}^2\,k_{\text{time}}(\cdot,\cdot)\bigr),
\\[0.2em]
g_{i,2}(\cdot) &\stackrel{\text{i.i.d.}}{\sim} \mathcal{GP}\bigl(0,\;\sigma_{g_2}^2\,k_x(\cdot,\cdot)\bigr),
\qquad i=1,\dots,m,
\end{align}
with all processes mutually independent and independent of the noise. $k_{\text{time}}$ is chosen among RBF and OU kernels on time,  while $k_x$ is an RBF kernel on covariates in its either in standard or in ARD form.

Let $K_{\text{time}} \in \mathbb{R}^{T\times T}$ be the covariance matrix with entries
$(K_\text{time})_{ts} = k_\text{time}(t,s)$. For each unit $i$, let
$K_{x,i} \in \mathbb{R}^{T\times T}$ have entries
$(K_{x,i})_{ts} = k_x(x_{it},x_{is})$. Similar to the structure of the time-only case, we derive
\[
\mathrm{Cov}\bigl(f_i(t),f_j(s)\bigr)
=
\begin{cases}
\sigma_\mu^2 k_{\text{time}}(t,s),
&
i \neq j, \\[0.3em]
\bigl(\sigma_\mu^2 + \sigma_{g_1}^2\bigr) k_{\text{time}}(t,s)
+ \sigma_{g_2}^2 k_x(x_{it},x_{is}),
&
i = j.
\end{cases}
\]
Thus the covariance of the stacked latent vector
$\mathbf{f} = (\mathbf{f}_1^\top,\dots,\mathbf{f}_m^\top)^\top$ takes the block form
\begin{equation}
\mathrm{Cov}(\mathbf{f})
=
\underbrace{(\mathbf{1}_m \mathbf{1}_m^\top) \otimes \bigl(\sigma_\mu^2 K_{\text{time}}\bigr)}_{\text{shared mean}}
\;+\;
\underbrace{I_m \otimes \bigl(\sigma_{g_1}^2 K_{\text{time}}\bigr)}_{\text{time deviations}}
\;+\;
\underbrace{\mathrm{diag}\bigl(\sigma_{g_2}^2 K_{x,1},\dots,\sigma_{g_2}^2 K_{x,m}\bigr)}_{\text{covariate deviations}}.
\label{eq:exch-Kf-covariates}
\end{equation}

The off-diagonal blocks are still given by the shared time kernel, but the diagonal blocks now differ across units, because of the unit-specific differences in $K_{x,i}$. In this setting, we preserve the structure of the shared mean, but we relax the exact exchangeability, as the distribution of unit-specific functions is no longer i.i.d. The marginal distribution of $\mathbf{y}$ is obtained as in \eqref{eq:exch-marginal-y} by adding the noise covariance $\Omega$ to $\mathrm{Cov}(\mathbf{f})$ from \eqref{eq:exch-Kf-covariates}.

We now extend the specification to allow different sized time grids for different units. In this case we lose the convenient Kronecker representation in \eqref{eq:exch-Kf-covariates}, but a comprehensive unit-specific representation is still possible. For $\{f_i(t)\}$, the covariance term has the general form
\[
\mathrm{Cov}\bigl(f_i(t),f_j(s)\bigr)
= \sigma_\mu^2\,k_{\text{time}}(t,s)
\;+\;
\mathbb{I}\{i=j\}
\Bigl[
  \sigma_{g_1}^2\,k_{\text{time}}(t,s)
  + \sigma_{g_2}^2\,k_x\bigl(x_{it},x_{is}\bigr)
\Bigr],
\]
where $\mathbb{I}\{i=j\}$ is the indicator of $i=j$. More specifically, the
variance of a single latent function is
\[
\mathrm{Var}\bigl(f_i(t)\bigr)
=
\bigl(\sigma_\mu^2 + \sigma_{g_1}^2\bigr)\,k_{\text{time}}(t,t)
+ \sigma_{g_2}^2\,k_x\bigl(x_{it},x_{it}\bigr).
\]

We use the full set of observations of the non-treated units and the pre-intervention observations of the treated unit as the set of controls to fit to the GP model and estimate the hyperparameters by maximizing the log marginal likelihood. We then condition on the covariates of the treated unit, as well as the controls, to derive the posterior predictive distribution and estimate the counterfactual path for the treated unit $i^{\star}$.

The set of parameters to be estimated includes the variance components $(\sigma_\mu^2, \sigma_{g_1}^2, \sigma_{g_2}^2)$, the unit-specific noise variances $\{\omega_i^2\}_{i=1}^m$ and the lengthscale parameters $(\ell_{\text{time}}, \boldsymbol{\ell}_x)$, where $\boldsymbol{\ell}_x = (\ell_1,\dots,\ell_p)^\top$ is a vector of length-scales corresponding to each of the $p$ covariate dimensions used in the ARD specification. In the standard RBF kernel setting on the covariates, $\boldsymbol{\ell}_x$ reduces to a single parameter.

One of the main advantages of the multi-task exchangeable GPs, compared to other multi-task GP specifications, is their computational efficiency. Because the models have a simpler structure induced by the exchangeable specification and fixed loadings on the latent functions $\mu$ and $g$, there are fewer hyperparameters to estimate, leading to a more stable and computationally efficient methodology. In our application with staggered adoption, we further increase the computational efficiency and make it applicable to large datasets with a pool of untreated units. Moreover, the decomposition of the latent function into unit-specific and shared components yields interpretable results and convenient modeling structure, where units with noisy paths can be pooled more towards the common mean. This helps stabilize the counterfactual estimates.

\chapter{Staggered adoption design}
\label{ch:staggered-adoption}

Many empirical policy settings involve multiple treated units--often hundreds or thousands--observed in a common panel alongside a pool of never-treated control units. Treatment adoption is staggered: each treated unit $i \in \mathcal{I}_1$ receives the intervention at its own time $T_{0i}$, while units in $\mathcal{I}_0$ remain untreated throughout. In such environments, one of the key restrictions is a no-interference condition: treatment assigned to one unit does not affect outcomes of other units. Under this assumption, the potential outcome of unit $i$ depends only on its own treatment status and not on the treatment paths of other units, so we remain within the assumptions specified in Section~\ref{sec:assumptions}.

The exchangeable GP model is naturally suited to the staggered-adoption structure, because it is designed to pool information across units. In principle, one can fit the exchangeable multi-task GP to the entire panel, using all pre-treatment observations of the treated units and all observations of the never-treated units, and then produce counterfactual predictions for the untreated potential outcomes of all treated units simultaneously. This approach uses the full set of available untreated information and learns the shared latent structure jointly across units.

The practical difficulty is computational. In the case of exact GP, each evaluation of the log marginal likelihood and its derivatives requires Cholesky factorization of an $n\times n$ matrix, whre $n$ denotes the total number of training observations (all control-unit observations plus all pre-treatment treated-unit observations). This makes each optimization step expensive, especially for datasets with thousands or tens of thousands observed units.

To make the problem tractable, we adopt a one-unit-at-a-time framework based on sampling from the pool of never-treated units. For each treated unit $i \in \mathcal{I}_1$, we construct a smaller training set consisting of that unit's pre-treatment observations together with a random subset of $M$ untreated units sampled from $\mathcal{I}_0$. If $M$ is sufficiently large and the sampling is random, the selected controls approximate the full untreated distribution, while keeping the computational cost menageable. We then treat each such problem as a single-treated-unit counterfactual prediction task and repeat this procedure independently for each treated unit.

With this design, the relevant matrix dimension is reduced from $n$ to
\[
n_i \;=\; T_{0i} + M\,T,
\]
where $T$ is the number of time points available for each untreated unit, $M$ is the number of controls, and $T_{0i}$ is the number of pre-treatment observations for treated unit $i$. Consequently, the total computational cost of one full pass over all treated units drops from $\mathcal{O}(n^3)$ to $\mathcal{O}\!\left(\sum_{i\in\mathcal{I}_1} n_i^3\right)$ time, enabling us to fit exchangeable GP models repeatedly across many treated units in large staggered-adoption panels.

\chapter{Estimation and Implementation}
\label{ch:estimation}

We now describe how the exchangeable Gaussian process models of Section~\ref{ch:exchangeable-gp} are estimated and how they are used to construct counterfactual trajectories. The basic ingredients of the model are the collection of latent untreated outcome values, stacked in a vector $\mathbf{f}$, and the vector of hyperparameters $\theta$, which includes variance components, observation noise variances and different length-scale parameters. The hyperparameters in $\theta$ differ depending on the exchangeable model specification, as described in section ~\ref{ch:exchangeable-gp}.

Together they define a hierarchical specification of the form
\[
\mathbf{f}\mid\theta \sim \mathcal{N}\bigl(0,K_f(\theta)\bigr),
\qquad
\mathbf{y}\mid\mathbf{f},\theta \sim \mathcal{N}\bigl(\mathbf{f},\Omega(\theta)\bigr),
\]
where $\mathbf{y}$ stacks the untreated outcome observations used for fitting,
$K_f(\theta)$ is the exchangeable covariance matrix, specified in section~\ref{ch:exchangeable-gp}, and $\Omega(\theta)$ is a diagonal noise covariance matrix. Conditional on $\theta$ the latent vector $\mathbf{f}$ follows a multivariate normal prior, and conditional on $\mathbf{f}$ the data are generated from a Gaussian likelihood.

Because both the prior and the likelihood terms are Gaussian, the marginal likelihood
$p(\mathbf{y}\mid\theta)$ can be obtained in closed form by integrating out the
latent vector:
\[
\mathbf{y}\mid\theta \sim \mathcal{N}\bigl(0,\Sigma(\theta)\bigr),
\qquad
\Sigma(\theta) = K_f(\theta) + \Omega(\theta).
\]
This analytic marginal likelihood is the main tool for hyperparameter learning. We adopt a Type~II maximum likelihood (empirical Bayes) approach \citep{rasmussen2006} (In the empirical implementation we use the \texttt{GPflow} library for Python):
we choose a point estimate
\[
\hat{\theta}
=
\arg\max_{\theta} \log p(\mathbf{y}\mid\theta),
\]

After learning the hyperparameters $\hat{\theta}$, we treat them as fixed and turn to prediction. Let $\mathbf{y}$ be the full stacked vector of untreated potential outcomes of controls and the pre-treatment untreated potential outcomes of the treated unit, and let $Z$ denote the corresponding inputs, where each row of $Z$ contains the time index and, when relevant, the covariates of a given observation. Suppose we now wish to predict the vector of post-intervention potential outcomes under no treatment $\mathbf{y}^\star$ at a collection of inputs $Z^\star$ for a treated unit $i^\star$.

Under the GP model with Gaussian likelihood, the joint distribution of the non-treated outputs $\mathbf{y}$ (referred to as \emph{training outputs}) and the corresponding non-treated outputs of the treated unit, $\mathbf{y}^\star$ (referred to as \emph{prediction outputs}), is multivariate normal,
\[
\begin{bmatrix}
\mathbf{y} \\
\mathbf{y}^\star
\end{bmatrix}
\Bigm|\hat{\theta}
\sim
\mathcal{N}\!\left(
\mathbf{0},
\begin{bmatrix}
K(Z,Z;\hat{\theta}) + \Omega & K(Z,Z^\star;\hat{\theta}) \\
K(Z^\star,Z;\hat{\theta})    & K(Z^\star,Z^\star;\hat{\theta}) + \Omega^\star
\end{bmatrix}
\right),
\]
where $K(Z,Z;\hat{\theta})$ is the covariance matrix of the latent GP at the training inputs $Z$, $K(Z^\star,Z^\star;\hat{\theta})$ is the corresponding matrix at the prediction inputs $Z^\star$, and $K(Z,Z^\star;\hat{\theta})$ is the latent cross-covariance between training and prediction locations. The diagonal matrices $\Omega$ and $\Omega^\star$ collect the observation-noise variances at the training and prediction points, respectively. All of these matrices are determined by the exchangeable covariance structure and the hyperparameters $\hat{\theta}$.

Conditioning this joint Gaussian distribution on the observed data
$(Z,\mathbf{y})$ and prediction inputs $Z^\star$ yields the posterior predictive distribution
\[
\mathbf{y}^\star \mid Z,Z^\star,\mathbf{y},\hat{\theta}
\sim
\mathcal{N}\bigl(m^\star, V^\star\bigr),
\]
with mean vector and covariance matrix
\begin{align}
m^\star
&=
K(Z^\star,Z;\hat{\theta})\,
\bigl[K(Z,Z;\hat{\theta}) + \Omega\bigr]^{-1}\,\mathbf{y},
\label{eq:gp-pred-mean-simple}
\\[0.3em]
V^\star
&=
K(Z^\star,Z^\star;\hat{\theta}) + \Omega^\star
-
K(Z^\star,Z;\hat{\theta})\,
\bigl[K(Z,Z;\hat{\theta}) + \Omega\bigr]^{-1}\,
K(Z,Z^\star;\hat{\theta}).
\label{eq:gp-pred-var-simple}
\end{align}
The covariance $V^\star$ depends only on the $(Z,Z^\star)$ and on the hyperparameters $\hat{\theta}$; it is completely independent of the realized outcomes $\mathbf{y}$. By contrast, the predictive mean $m^\star$ is a linear combination of the observed outcomes, with weights determined by the covariance
structure and the noise level.

The predictive mean $m^\star$ therefore provides the counterfactual trajectory $\{y_{i^\star t}(0)\}$ for the treated unit, and $V^\star$ quantifies the associated uncertainty. These quantities form the basis for the causal effect estimates reported later in Section~\ref{ch:application}.

\chapter{Applications}
\label{ch:application}

In this section, we apply the exchangeable Gaussian process model presented in Section~\ref{ch:exchangeable-gp} to estimate treatment effects in two different settings.

The first dataset concerns Proposition~99 (Prop99), a tobacco control program implemented in California in 1988 aimed at reducing cigarette sales.

The second dataset comprises a large panel of Greek petrol stations, where audits were conducted at various times to improve the integrity and performance of the monitoring infrastructure.

The two applications differ fundamentally: the Prop99 setting involves a single treated unit with once-and-for-all treatment, whereas the Greek dataset features hundreds of treated and control units with staggered treatment times.
However, our procedure described in Section~\ref{ch:staggered-adoption} allows us to approach both problems from a single-treated-unit perspective.

\section{Application 1: Proposition~99 (Prop99)}
\label{sec:prop99}

\begin{figure}[htbp]
  \centering
  \includegraphics[width=0.9\textwidth]{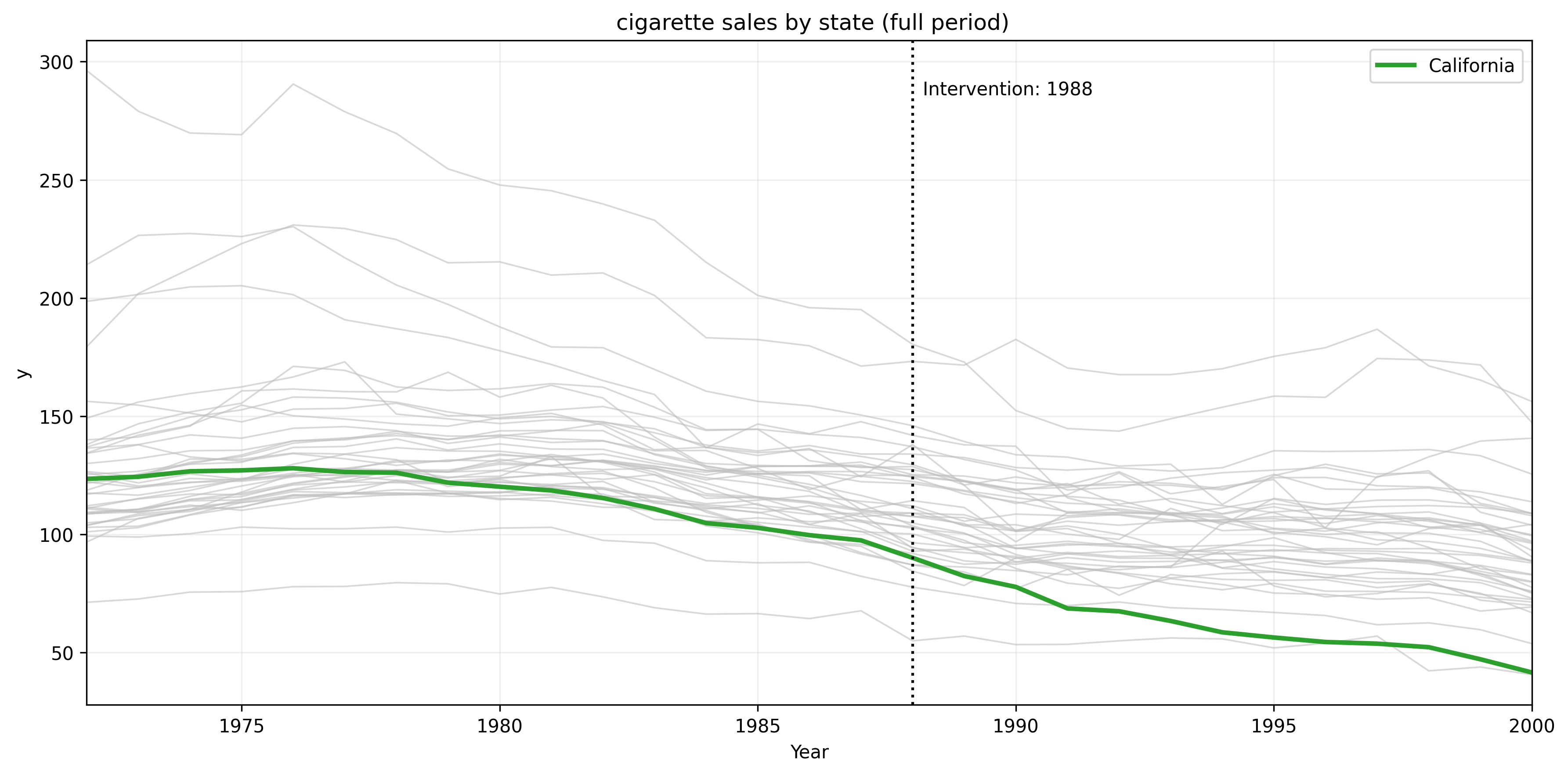}
  \caption{Per-capita cigarette sales in California and other U.S. states (1970--1999).}
  \label{fig:all-states-y}
\end{figure}

Our first applied case refers to the dataset used by \cite{AbadieDiamondHainmueller2010} in a synthetic control application, where a tobacco control program was introduced in California in 1988.
In the following years a sharper decline in per-capita cigarette sales is recorded in the state, compared to other states (see Figure~\ref{fig:all-states-y}). The main objective is to assess whether this post 1988 trajectory is attributed to the prop99 with the use of Exchangeable GP models. For that, we aim to construct a counterfactual trajectory, which will recover the sales reports, that would have been observed had the program not been implemented. By using the other states' results as controls, alongside the observed covariates, we construct the posterior predictive distribution of $y_{\text{CA},t}(0)$ for $t > 1988$ and derive pointwise and cumulative treatment effect estimates with credible intervals.

We first describe the dataset and its key variables, and we define the outcome and covariates used in the analysis. Next, we outline the validation strategy, introduce a benchmark comparison based on synthetic difference-in-differences (SDID), and describe how the counterfactual paths and average treatment effects are constructed from the GP posterior predictive distribution. Finally, we present the results of the validation experiment and the estimated effects of Prop99 on cigarette sales.

\subsection{Data description}
\label{subsec:data-eval-design-ca}

Our empirical application uses a state-level panel dataset. Units are U.S.\ states observed annually. We index
states by $i=1,\dots,m$ and years by $t \in \{1970,\dots,1999\}$.

The outcome is per-capita cigarette sales. We use California as the single treated unit
($i^\star$) and 38 U.S.\ states as controls. These states are chosen because they did not implement similar policy interventions.

The dataset includes a set of time-varying covariates: log per-capita income, per-capita beer consumption, and the share of the population aged 15--24. In addition, we use all-states per-capita values as global covariates. These global covariates are used as additional inputs for the shared function $\mu$ (specified in Section~\ref{ch:exchangeable-gp}).

Proposition~99 was implemented in California in 1988, and the changes took effect in 1989. Hence, we treat 1970--1988 as the pre-intervention period and 1989--1999 as the post-intervention period.

\subsection{Validation and treatment effect estimation design.}
\label{sec:validation_design}

Before using the GP models for treatment effect estimation, we need to verify whether they can reconstruct realistic untreated trajectories.
To first assess predictive accuracy, we only use observations from years $1970$--$1988$, where no unit is treated and the true outcomes are observed.

We then choose a ``fake treatment'' year $t_1$ within this pre-intervention window so that the fraction of post ``fake-treatment'' observations in the validation setup matches the fraction of post-treatment observations in the actual application. In our setting, this yields $t_1 = 1981$. Thus, years $1970$--$1981$ constitute the (fake) pre-treatment period and years $1982$--$1988$ constitute the (fake) post-treatment period.

\begin{figure}[H]
  \centering
  \includegraphics[width=0.95\textwidth]{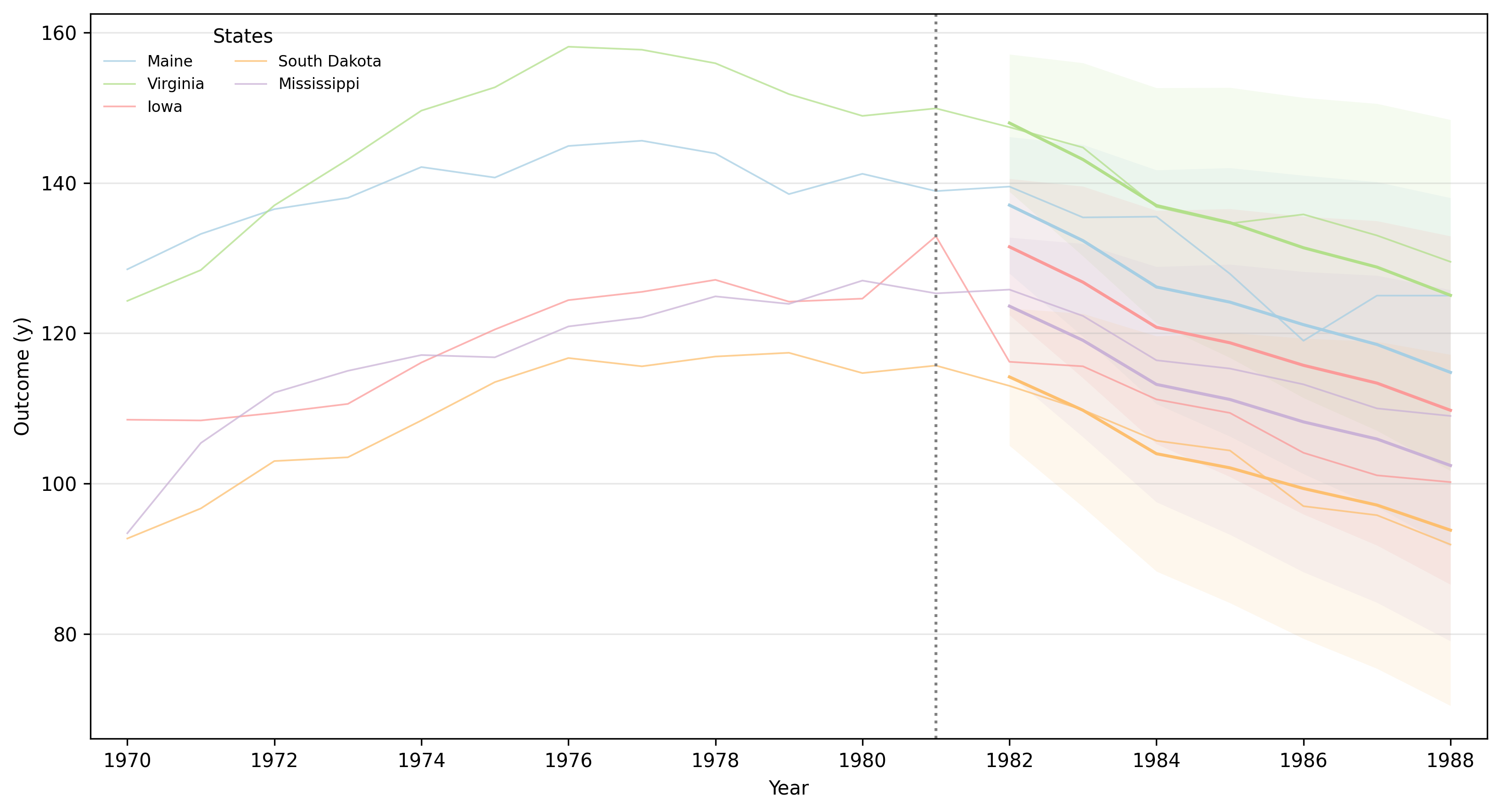}
  \caption{Example run of the GP--OU--time--covariates model for five randomly chosen pseudo-treated states in the pre-treatment validation step. The bold lines show the posterior predictive means, and the transparent bands show the corresponding 95\% prediction intervals.}
  \label{fig:pre-treatment-validation-ou-time-covs-5-states}
\end{figure}

We implement a leave-one-state-out design. For each state $i$ in turn, we treat it as the pseudo-treated unit. We fit the exchangeable GP model using the outcomes of all other states over years $1970$--$1988$ as controls and the pseudo-treated state's outcomes in the fake pre-treatment period ($1970$--$1981$). Conditioning on the full set of covariates, we predict the pseudo-treated state's outcomes for the post ``fake-treatment'' years ($1982$--$1988$) and compare them to the realized values (see Figure~\ref{fig:pre-treatment-validation-ou-time-covs-5-states} for an example run of the GP-OU-time-covariates model for five randomly chosen pseudo-treated states).

To evaluate the predictive performance of the different model specifications, we use several accuracy metrics. For each pseudo-treated state, we compute the mean absolute percentage error (MAPE) over its validation predictions, the root mean squared error (RMSE), and the predictive bias. We report the average values. We also compute the percentage of predicted points that fall within the $95\%$ prediction intervals, as well as the average width of these intervals. To assess the computational cost of the methods, we report the average per-unit optimization time (in seconds). This yields Table~\ref{tab:validation-summary-ca}.

We evaluate multiple exchangeable GP kernel specifications under this leave-one-state-out validation design. We consider the following specifications: (i) GP--RBF--time, which uses a squared-exponential (RBF) kernel over time to capture smooth nonlinear trends; (ii) GP--OU--time, which uses an Ornstein--Uhlenbeck kernel to allow for less smooth dynamics; (iii) GP--RBF--time--covariates, which combines the GP--RBF--time kernel with covariates; and (iv) GP--OU--time--covariates, which combines the OU time kernel with covariates.

\paragraph{Benchmark comparison: synthetic difference-in-differences}

As a benchmark, we also compare our exchangeable GP approach to synthetic difference-in-differences (SDID) \citep{arkhangelsky2021sdid}. SDID is a frequentist method, but it is widely used as a modern benchmark in comparative case studies and policy evaluation.
SDID can be viewed as a weighted version of a TWFE model.

To match the strategy used for our GP models, we implement SDID in a leave-one-state-out fashion. Using only the pre-treatment data, for each state $i$ in turn, we treat it as the pseudo-treated unit, learn the parameters, fit the weighted TWFE model using all observations unaffected by the ``fake treatment'' (all other states plus the pseudo-treated state's pre-intervention outcomes), and then predict the counterfactual outcomes over the validation (fake post-treatment) window.

As in the GP case, we report the same set of predictive summary quantities (MAPE/RMSE, bias, prediction-interval coverage/width, and per-unit optimization time) and summarize the implied treatment effects over time and in aggregate.

\paragraph{Treatment effect estimation design.}
Once the validation step is complete and the best-performing model is selected, we return to the actual intervention date (1988) and estimate the causal effect of Prop~99 for California. We fit the model using outcomes for the control states and California's pre-1988 outcomes, alongside the covariates, and then predict California's counterfactual outcomes for the years $1989$--$1999$.

Given realized post-intervention outcomes $y_{\text{CA},t}$ and their predicted counterfactual counterparts $\hat y_{\text{CA},t}(0)$, we calculate the time-specific treatment effect as $\hat\delta_{\text{CA},t} = y_{\text{CA},t} - \hat y_{\text{CA},t}(0)$. We report cumulative and average post-intervention effects, together with uncertainty intervals (see Table~\ref{tab:att-post-comparison} and Figure~\ref{fig:ca-post-compare-ou-synthdid}).

\subsection{Results}
\label{subsec:results}

\paragraph{Validation of model performance.}
We first present the results of the validation step described in Section~\ref{sec:validation_design}. In this exercise, we set a ``fake treatment'' year to 1981, predict outcomes for the post-``fake treatment'' window (1982--1988) for all states, and compare the predictions with the observed outcomes.

Table~\ref{tab:validation-summary-ca} summarizes the average predictive accuracy and uncertainty metrics across all pseudo-treated units and validation years for the exchangeable GP specifications and the benchmark synthetic difference-in-differences (SynthDiD) method.

Overall,the SynthDiD benchmark attains the lowest prediction error (RMSE and MAPE). Among the exchangeable GP models, the OU specification over time with covariates (OU--time--covariates) delivers the best average point-prediction accuracy (lowest RMSE, MAPE, and bias), while achieving approximately $95\%$ interval coverage. The RBF time-only model performs the worst, with noticeably larger RMSE and more negative bias, indicating that time-only smoothness assumptions are insufficient to reconstruct untreated trajectories in this setting. Adding covariates improves predictive accuracy for both OU and RBF specifications.

\begin{table}[htbp]
  \centering
  \caption{Prop99 Validation summary: averaged predictive accuracy, intraclass correlation, and runtime across methods (fake treatment at 1981; predictions for 1982--1988).}
  \label{tab:validation-summary-ca}
  \scriptsize
  \begin{tabular}{lccccccc}
    \toprule
    Method & $\rho$ & MAPE & RMSE & Bias & Coverage (95\%) & Avg. PI width & Opt. time (s) \\
    \midrule
    OU (Time)                        & 0.869 & 0.051 & 7.255 & $-0.829$ & 0.956 & 34.35 & 3.121 \\
    OU (Time + Covariates)           & 0.839 & 0.051 & 7.251 & $-0.348$ & 0.956 & 34.34 & 6.156 \\
    RBF (Time)                       & 0.599 & 0.070 & 9.302 & $-1.551$ & 0.927 & 40.71 & 2.992 \\
    RBF (Time + Covariates)          & 0.715 & 0.053 & 7.458 & $-0.369$ & 0.971 & 36.65 & 11.189 \\
    SynthDiD                         & --    & 0.047 & 6.884 & 0.436 & - & - & 0.164 \\
    \bottomrule
  \end{tabular}
  \begin{tablenotes}
    \footnotesize
    \item \textit{Note:} Mean absolute percentage error (MAPE), root mean squared error (RMSE), and bias are averaged over all states and validation years. intraclass correlation: $\rho = {\sigma_g^2}/{(\sigma_\mu^2 + \sigma_g^2)}$, takes values between $0$ and $1$ and measures the proportion of total variance that is unit specific.
  \end{tablenotes}
\end{table}

\begin{figure}[H]
  \centering
  \includegraphics[width=0.95\textwidth]{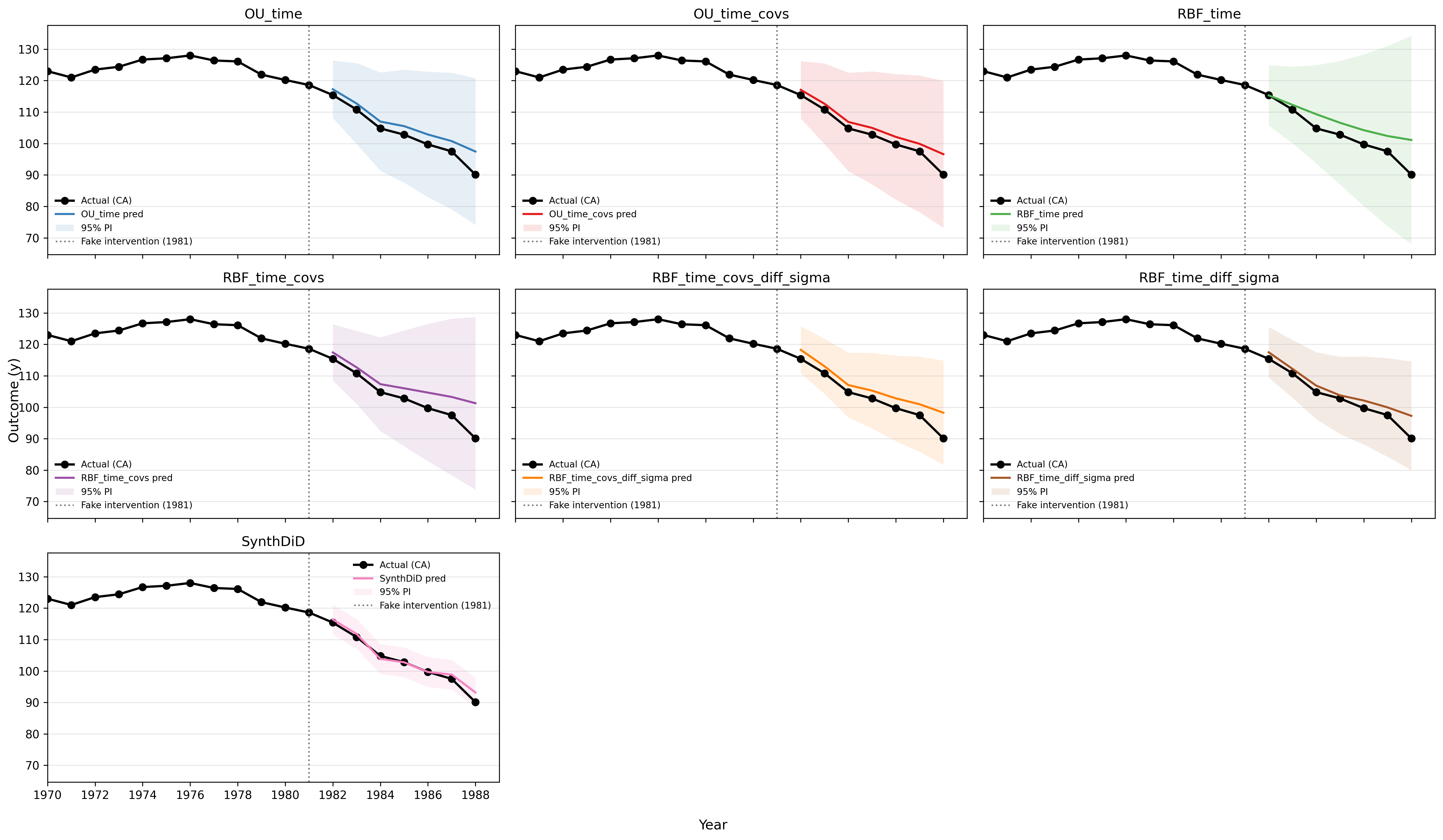}
  \caption{Validation example for California with a fake treatment in 1981: observed outcomes and posterior predictive counterfactual path over 1982--1988.}
  \label{fig:validation-ca-fake-1981}
\end{figure}

\begin{figure}[H]
  \centering
  \includegraphics[width=0.95\textwidth]{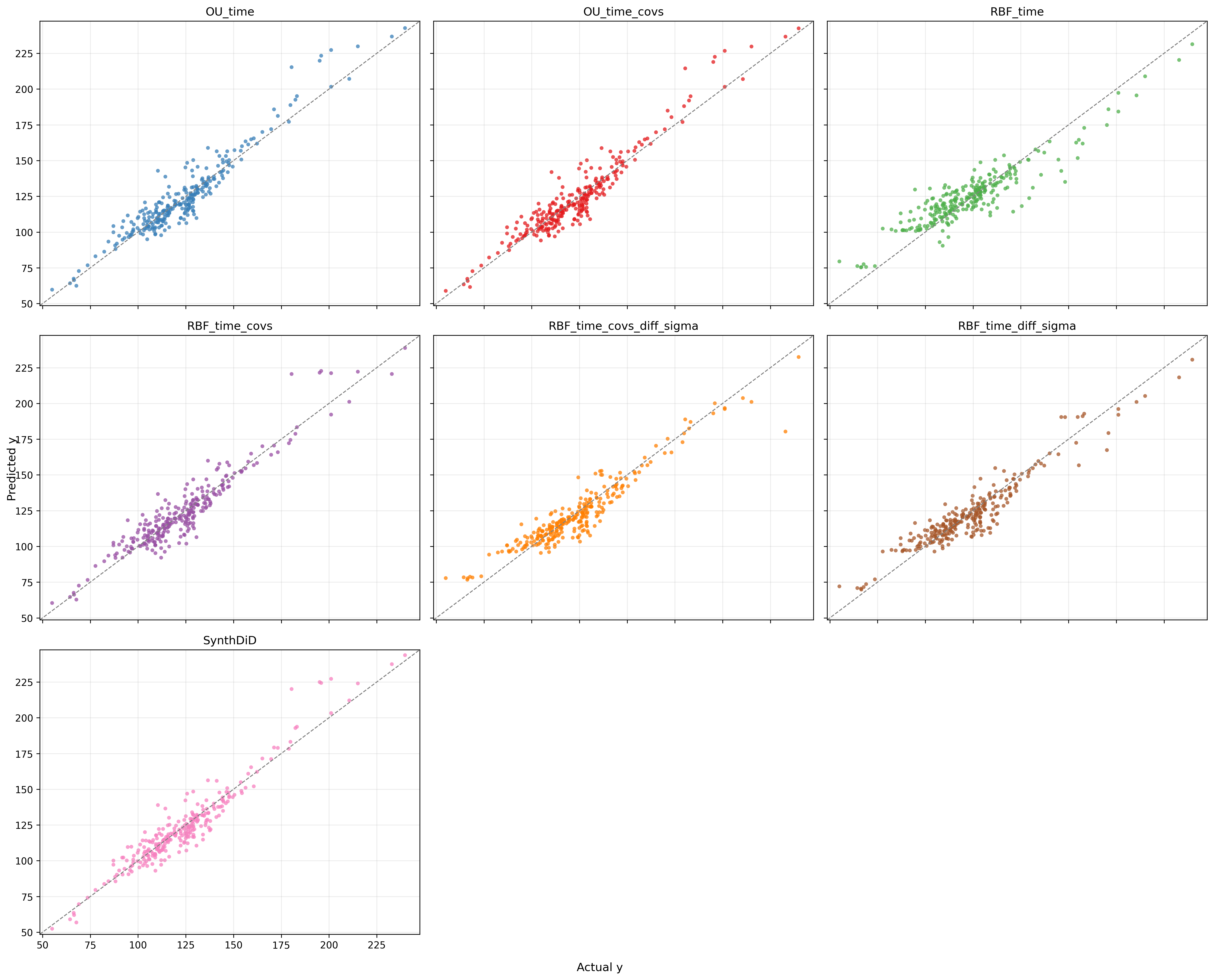}
  \caption{Predicted versus observed outcomes in the validation exercise, by method.}
  \label{fig:validation-scatter-by-method}
\end{figure}

Figure~\ref{fig:validation-ca-fake-1981} illustrates the validation results for California under the ``fake treatment'' at 1981, showing the actual pre-intervention path and the predicted counterfactual trajectory for the validation period. The assessment provides a clear indication that the SynthDiD method implies higher predictive accuracy for California. To visually assess all-states predictive accuracy of the methods, Figure~\ref{fig:validation-scatter-by-method} reports predicted-versus-observed comparisons for all.

\FloatBarrier

\paragraph{Estimated causal effects.}
Based on the validation results reported above, we select the best-performing exchangeable GP specification, OU--time--covariates, for the final empirical analysis of Proposition~99.
We then fit this model on the full pre-intervention period for California (1970--1988), together with the full data for the control states, and predict California's counterfactual outcomes for the post-intervention period (1989--1999). As a benchmark, we also report results from SynthDiD; this method is included purely as a comparison with a widely used frequentist approach.

Table~\ref{tab:att-post-comparison} summarizes the aggregated causal estimands for California over the 11 post-intervention years (1989--1999). Both methods indicate a negative average treatment effect (a reduction in cigarette sales) and a negative cumulative effect over the post-intervention period. However, the magnitudes differ: the exchangeable model suggests a larger reduction than SynthDiD.

\begin{table}[H]
  \centering
  \caption{Aggregated post-intervention effects for California (1989--1999): exchangeable GP (OU--time--covariates) versus SynthDiD benchmark.}
  \label{tab:att-post-comparison}
  \small
  \resizebox{\linewidth}{!}{%
  \begin{tabular}{lcccc}
    \toprule
    Method & $\tau_{i^\star}$ & 95\% CI & $\text{C}_{i^\star}$ & 95\% CI \\
    \midrule
    OU--time--covariates & $-19.588$ & $[-26.021,\,-13.155]$ & $-215.469$ & $[-286.233,\,-144.705]$ \\
    SynthDiD (benchmark) & $-14.386$ & $[-15.884,\,-12.888]$ & $-158.247$ & $[-174.729,\,-141.764]$ \\
    \bottomrule
  \end{tabular}%
  }
  \begin{tablenotes}
    \footnotesize
    \item \textit{Note:} ``$\tau_{i^\star}$'' is the average treatment effect over 1989--1999 (11 years) and ``$\text{C}_{i^\star}$'' is the corresponding cumulative effect. SynthDiD is reported as a benchmark comparison.
  \end{tablenotes}
\end{table}

\begin{figure}[H]
  \centering
  \includegraphics[width=0.95\textwidth]{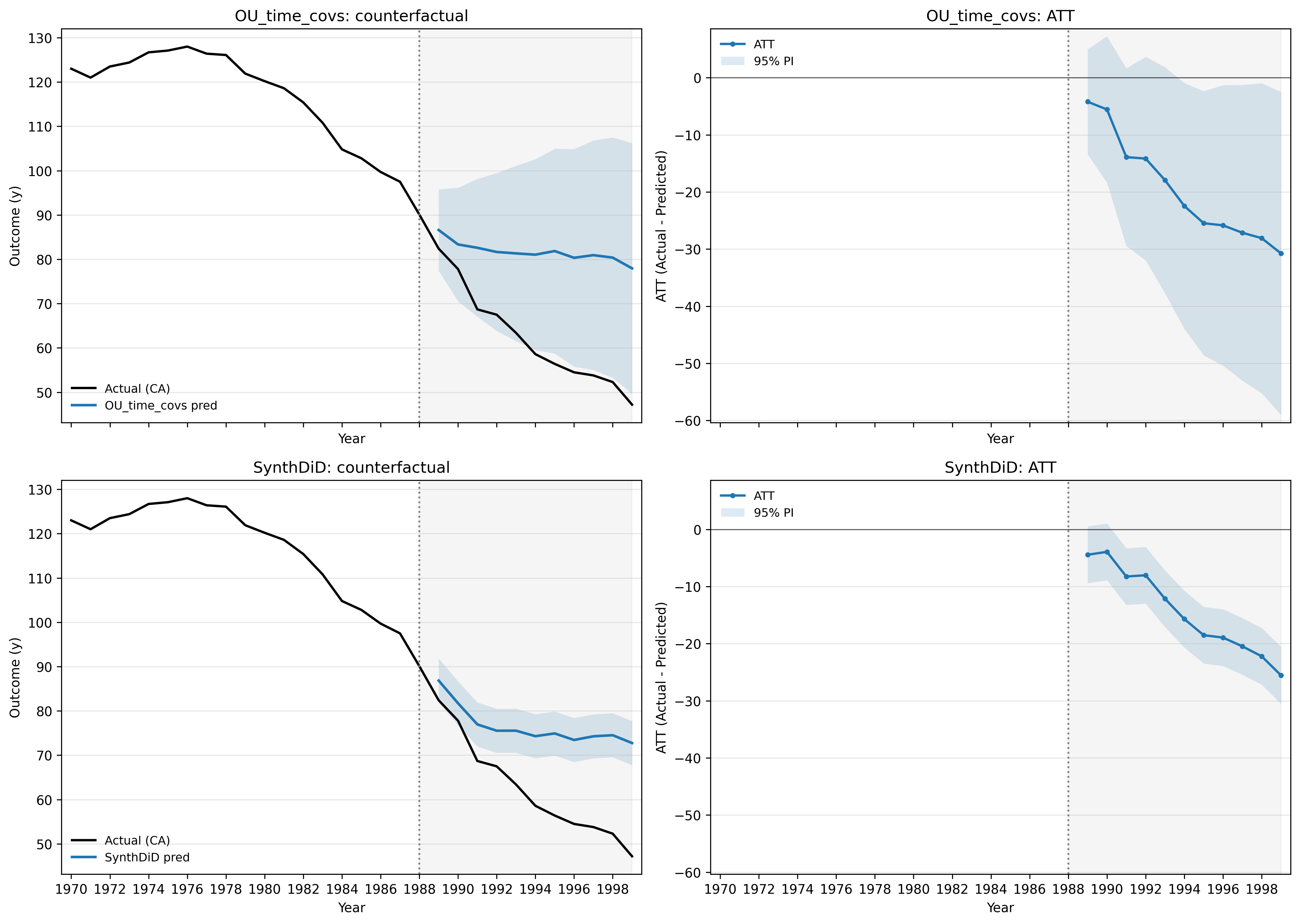}
  \caption{California post-intervention comparison (1989--1999): exchangeable GP (OU--time--covariates) vs. SynthDiD benchmark.}
  \label{fig:ca-post-compare-ou-synthdid}
\end{figure}

Figure~\ref{fig:ca-post-compare-ou-synthdid} plots the treated outcome and the two counterfactual trajectories for California in the post-intervention period, together with the implied treatment effects.Shaded regions show 95\% prediction intervals. The exchangeable GP band is posterior predictive, while the SynthDID band is a frequentist predictive interval that propagates uncertainty from estimated parameters/weights and idiosyncratic error.

Taken together, both approaches suggest that Proposition~99 is associated with a reduction in cigarette consumption in California relative to its estimated counterfactual path.

\FloatBarrier
\section{Application 2: Staggered-adoption audit on Greek petrol stations}
\label{sec:appl2}

In this section, we apply the exchangeable Gaussian process framework, presented in Section~\ref{ch:exchangeable-gp}, to a large panel dataset of Greek petrol stations with staggered adoption. Each fuel station was observed weekly throughout 2023. The intervention of interest is an IAPR audit that took place in a specific week for a subset of stations, such that stations switched from ``not yet audited'' to ``audited'' at different points in time. Our aim is to assess the effect of audits on measured weekly sales volume.

We first present the dataset. We then describe the validation procedure used to assess the predictive accuracy of the exchangeable GP models in the staggered-adoption setting. Next, we outline our approach for predicting the counterfactual paths of the treated units and estimating average treatment effects. Finally, we present the results from both the validation and treatment effect estimation steps using the relevant graphs and tables. We keep the main aggregated tables and figures in the body of the section, and report secondary results in Appendix~\ref{app:supplement}.

\subsection{Data description}
\label{subsec:data-appl2}

A panel dataset of approximately $2{,}500$ Greek petrol stations is used for the application. The data are observed weekly over roughly one year ($T = 52$ weeks). Each station is indexed by $i=1,\dots,m$, and weeks are indexed by $t=1,\dots,T$. For each unit $i$, we observe weekly sales $y_{it}$ as the outcome variable and two covariates, $x_{1,it}$ and $x_{2,it}$, which summarize the number of alerts generated by the inflow--outflow monitoring system. Specifically, these covariates aggregate (i) technical/operational alerts (e.g., sensor or dispenser malfunctions, communication or data-transmission failures, missing or inconsistent readings) and (ii) reconciliation/compliance alerts (e.g., fuel-balance deviations exceeding predefined tolerance thresholds).

Among the observed units (i.e., petrol stations), approximately $1{,}500$ received the treatment (i.e., were audited) at some point in time, while the rest were never treated. Each treated unit receives its once-and-for-all treatment in a single week between weeks $23$ and $52$. In the final cleaned sample---after removing observations with missing values and extreme outliers and balancing the panel---we work with a subset of $864$ treated units and $426$ untreated units.

For the validation step, we use all observations for the untreated units and only the pre-treatment observations for the treated units. For the final treatment effect evaluation, we use the full dataset for both treated and untreated units.

\subsection{Validation and treatment effect estimation design.}
\label{sec:validation_design_appl2}

To assess the predictive accuracy of the exchangeable model, we design a validation procedure that mimics the final causal application as closely as possible while using only pre-intervention data, for which the true outcomes are known.

We conduct the validation procedure on a randomly chosen subset of $300$ treated units. For each treated unit $i$ from the subset, we proceed as follows. We choose a ``fake treatment'' time strictly before the actual intervention period so that it splits the pre-treatment dataset, leaving one third of the observations in the post-``fake treatment'' period. For example, for a unit treated at week $30$, we retain only its pre-treatment observations and choose the ``fake treatment'' time to be week $20$. This splits the data into two parts: pre-``fake treatment'' observations for weeks $1$ to $20$ and post-``fake treatment'' observations for weeks $21$ to $30$. This step is intended to avoid predicting overly long intervals for some units, which could bias the aggregated results, especially for model specifications without covariates.

Model fitting and posterior prediction are conducted in the same manner as described in Section~\ref{ch:staggered-adoption}. In each experiment, we assume there is only a single treated unit and that the rest are untreated. To achieve this, for each treated unit $i$, we randomly choose twenty untreated units from the full pool of untreated units as controls. We fit this smaller dataset using the exchangeable Gaussian process model and predict the counterfactual path. During the validation step, we conduct this procedure for each model specification to derive aggregated prediction-accuracy metrics. This allows us to compare the six exchangeable GP specifications and choose the best-performing one for the final treatment-effect analysis.

To evaluate the predictive performance of the model specifications, we use several accuracy metrics. For each treated unit, we compute the root mean squared error (RMSE) over its validation predictions and the predictive bias, and report the average values. We also report the percentage of predicted points that fall within the $95\%$ posterior predictive intervals, as well as the average width of these intervals. This yields Table~\ref{tab:validation-summary-petrol}. Additionally, we report the same metrics by time point and by predictive horizon (reported only for the GP--RBF--time specification) to verify whether predictive accuracy declines with the horizon or varies across time points. These results are presented in Appendix~\ref{app:supplement} in Table~\ref{tab:horizon-rbf-time-covs}.

We evaluate predictive performance in the validation step using six exchangeable GP model specifications: (i) RBF--time--covariates, which uses an RBF kernel over time together with covariates; (ii) RBF--time--covariates--ARD, which is the same model but with automatic relevance determination (ARD) over the covariate dimensions; (iii) RBF--time, which uses an RBF kernel over time only; (iv) OU--time, which uses an Ornstein--Uhlenbeck (OU) kernel over time only; (v) OU--time + RBF--covariates, which combines an OU kernel over time with an RBF kernel over covariates; and (vi) OU--time + RBF--covariates--ARD, which adds ARD to the covariate kernel in the previous specification.

Once the validation step is complete and the best-performing model is selected, we return to the actual intervention dates and estimate the causal effect of the policy intervention on the outcomes of the treated units. Using the model with the highest predictive accuracy, we follow the same methodology to predict the counterfactual paths of the treated units, with a few distinctions. Most importantly, we now predict the counterfactual paths for all treated units, which corresponds to $864$ independent model runs. For each run, as in the validation case, we use a randomly chosen subset of $20$ untreated units as controls. Because units have varying numbers of post-treatment observations, we restrict the maximum post-intervention prediction horizon to $50\%$ of the treated unit's pre-intervention sample size. This means that, if a unit is treated at time $21$, we predict its counterfactual path only up to time $30$. This restriction helps avoid overly long prediction horizons, which might reduce predictive accuracy and skew the results.

To assess the causal effect of the treatment on the treated units, we present both the average, aggregate metrics and their corresponding uncertainty quantification. Given realized post-intervention outcomes $y_{it}$ and their predicted counterparts $\hat y_{it}(0)$, we calculate the unit-specific treatment effect at time $t$: $\hat\delta_{it} \;=\; y_{it} - \hat y_{it}(0)$. For each calendar week $t$, we take the average of the treatment effects over all treated units observed at that time. This yields the time specific average treatment effect on the treated units ($ATT$): $\widehat{\text{ATT}}_t \;=\; \frac{1}{n_t} \sum_{i \in \mathcal{N}_t} \hat\delta_{it}$, where $\mathcal{N}_t$ is the set of treated units with available post-intervention observations at $t$ and $n_t = |\mathcal{N}_t|$.
We also report the cumulative effect over the post-intervention window $\mathcal{T}_1$: $\widehat{C} \;=\; \sum_{t \in \mathcal{T}_1} \widehat{\text{ATT}}_t$, and the average per-period effect $\bar{\tau} \;=\; \frac{1}{|\mathcal{T}_1|} \sum_{t \in \mathcal{T}_1} \widehat{\text{ATT}}_t$. Additionally we report the 95\% credible intervals for $\widehat{\text{ATT}}_t$, $C$ and $\bar{\tau}$. This analysis corresponds to the results presented in table~\ref{tab:att-comparison} and the figure~\ref{fig:cumulative-att}, as well as table \ref{tab:att-time-rbf-time-covs}.

\subsection{Results}
\label{sec:results}

\paragraph{Validation of model performance.}
We first present the results of the validation step of the experiment, where we compare the predictive performence of the six GP exchangeable model specifications listed above. Table~\ref{tab:validation-summary-petrol} summarizes the average prediction accuracy metrics across all treated units and timepoints.

The results demonstrate that the RBF specifications including covariates achieve substantially lower RMSE and bias compared to the other specifications. The addition of covariate parameters noticably improves the predictive accuracy of the exchangeable models. Furthermore, all of the variants report well-calibrated coverage results close to the nominal 95\%,. Across all specifications, the estimated bias is negative.

\begin{table}[htbp]
  \centering
  \caption{Averaged prediction accuracy metrics for outcome $y$ across six exchangeable GP specifications during the validation period.}
  \label{tab:validation-summary-petrol}
  \small
  \setlength{\tabcolsep}{4pt}
  \resizebox{\textwidth}{!}{%
  \begin{tabular}{lccccc}
    \toprule
    Model Specification & RMSE & Bias & Coverage (\%) & PI Width & Avg conv. time (sec) \\
    \midrule
    RBF (Time + Covariates, ARD) & 0.267 & $-0.063$ & 94.7 & 1.41 & 16.81 \\
    RBF (Time + Covariates)      & 0.256 & $-0.050$ & 95.2 & 1.41 & 14.96 \\
    RBF (Time only)              & 0.608 & $-0.242$ & 87.6 & 2.31 & 10.36 \\
    OU (Time only)               & 0.379 & $-0.109$ & 91.9 & 1.68 & 10.65 \\
    OU (Time) + RBF (Covariates) & 0.358 & $-0.103$ & 91.8 & 1.54 & 14.67 \\
    OU (Time) + RBF (Covariates, ARD) & 0.358 & $-0.100$ & 92.2 & 1.55 & 16.27 \\
    \bottomrule
  \end{tabular}
  }%
  \begin{tablenotes}
    \footnotesize
    \item \textit{Note:} Metrics are averaged over all treated units and prediction periods. Coverage refers to the percentage of predicted values falling within the 95\% predictive interval.
  \end{tablenotes}
\end{table}

\begin{figure}[htbp]
  \centering
  \includegraphics[width=0.8\textwidth]{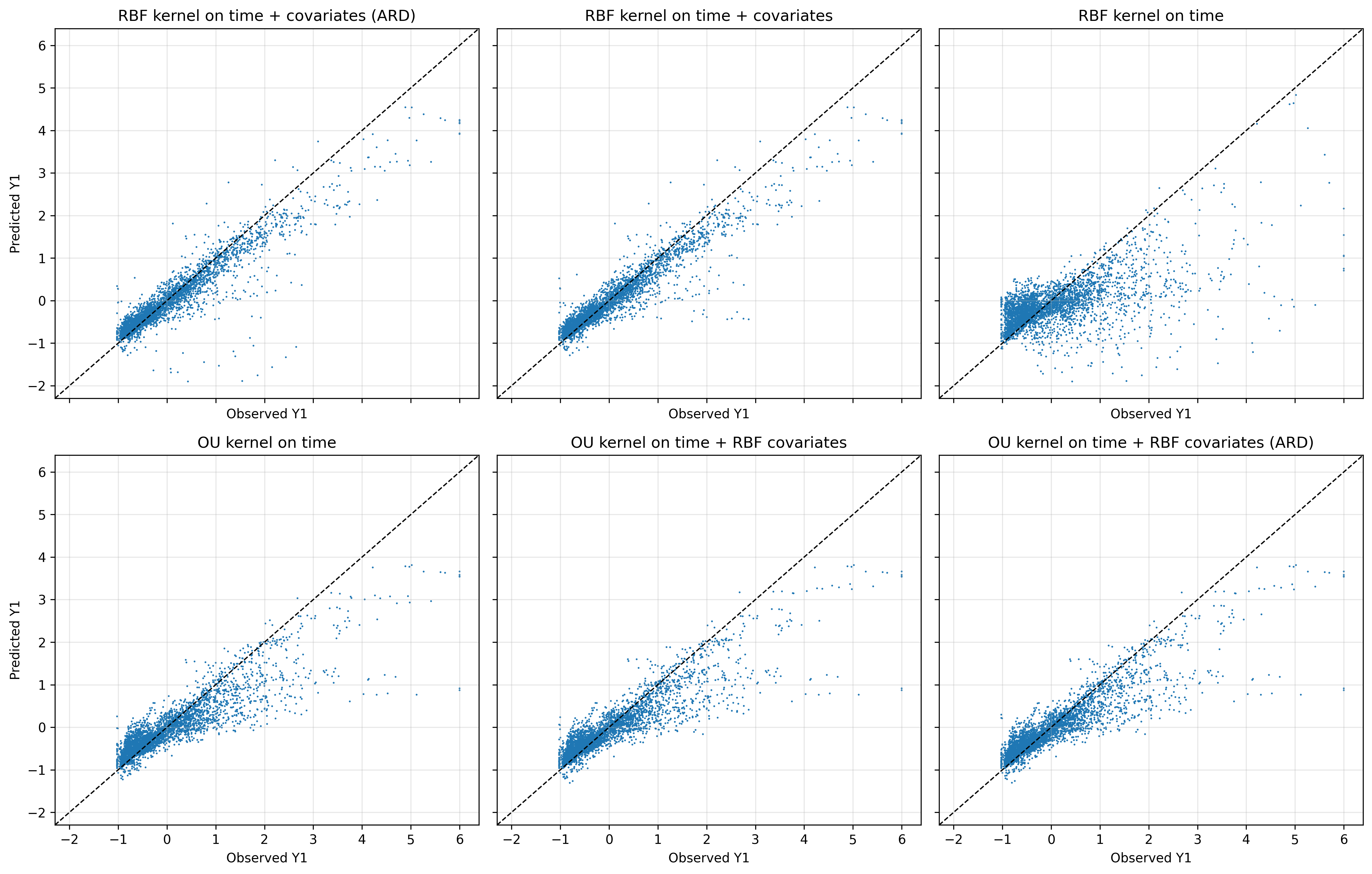}
  \caption{Scatter plot of predicted ($\hat y(0)$) versus observed outcomes ($y(0)$) for three model specifications. The red dashed line represents perfect prediction.}
  \label{fig:pred-vs-obs}
\end{figure}

To visually assess the goodness of fit for the best-performing model (RBF (Time + Covariates)), Figure~\ref{fig:pred-vs-obs} plots the predicted versus observed values for the outcome variable $y$. The points cluster tightly around the 45-degree line, indicating that the model captures the variation in $y$ without systematic deviation.

While Table~\ref{tab:validation-summary-petrol} provides aggregate metrics, predictive performance varies by prediction horizon and calendar time. Detailed horizon and time specific accuracy metrics are reported in Appendix~\ref{app:rbf-time-covs-results}. Additionally, representative posterior predictive trajectories for individual units can be found in Appendix~\ref{app:example-runs}. Based on the validation results, we proceed with the RBF--time-covs specification for the final causal analysis.

\paragraph{Estimated causal effects.}
To interpret the estimated treatment effects, we consider results from both the validation experiment and the final application. We report the ATT for the validation data to separate systematic features driven by model performance from actual treatment effects. If the policy had a substantial causal impact, we would expect the ATT estimates from the real intervention to differ noticeably from the validation estimates.

Table~\ref{tab:att-comparison} summarizes the aggregated causal estimands for both the validation and final setups. For the final empirical application, the model estimates a positive total cumulative ATT. However, neither the validation nor the final credible interval includes zero. Given that Table~\ref{tab:validation-summary-petrol} shows negative prediction bias across model specifications, it is natural for this to translate into a positive bias in the estimated ATT, even in the validation setting where no true treatment effect is present.

\begin{table}[htbp]
  \centering
  \caption{Comparison of aggregated Average Treatment Effects (ATT) between the Validation experiment and the Final (Real) application.}
  \label{tab:att-comparison}
  \small
  \begin{tabular}{lcccc}
    \toprule
    & \multicolumn{2}{c}{\textbf{Total Cumulative ATT}} & \multicolumn{2}{c}{\textbf{Average Weekly ATT}} \\
    \cmidrule(lr){2-3} \cmidrule(lr){4-5}
    Experiment & Estimate & 95\% CI & Estimate & 95\% CI \\
    \midrule
    Validation & 0.830 & $[0.442, 1.219]$ & 0.022 & $[0.012, 0.033]$ \\
    Final Application (Real) & 1.187 & $[0.975, 1.398]$ & 0.041 & $[0.034, 0.048]$ \\
    \bottomrule
  \end{tabular}
  \begin{tablenotes}
    \footnotesize
    \item \textit{Note:} The Validation period spans 37 weeks (fake treatment), while the Final Application period spans 29 weeks (actual intervention).
  \end{tablenotes}
\end{table}

\begin{figure}[htbp]
  \centering
  \includegraphics[width=0.9\textwidth]{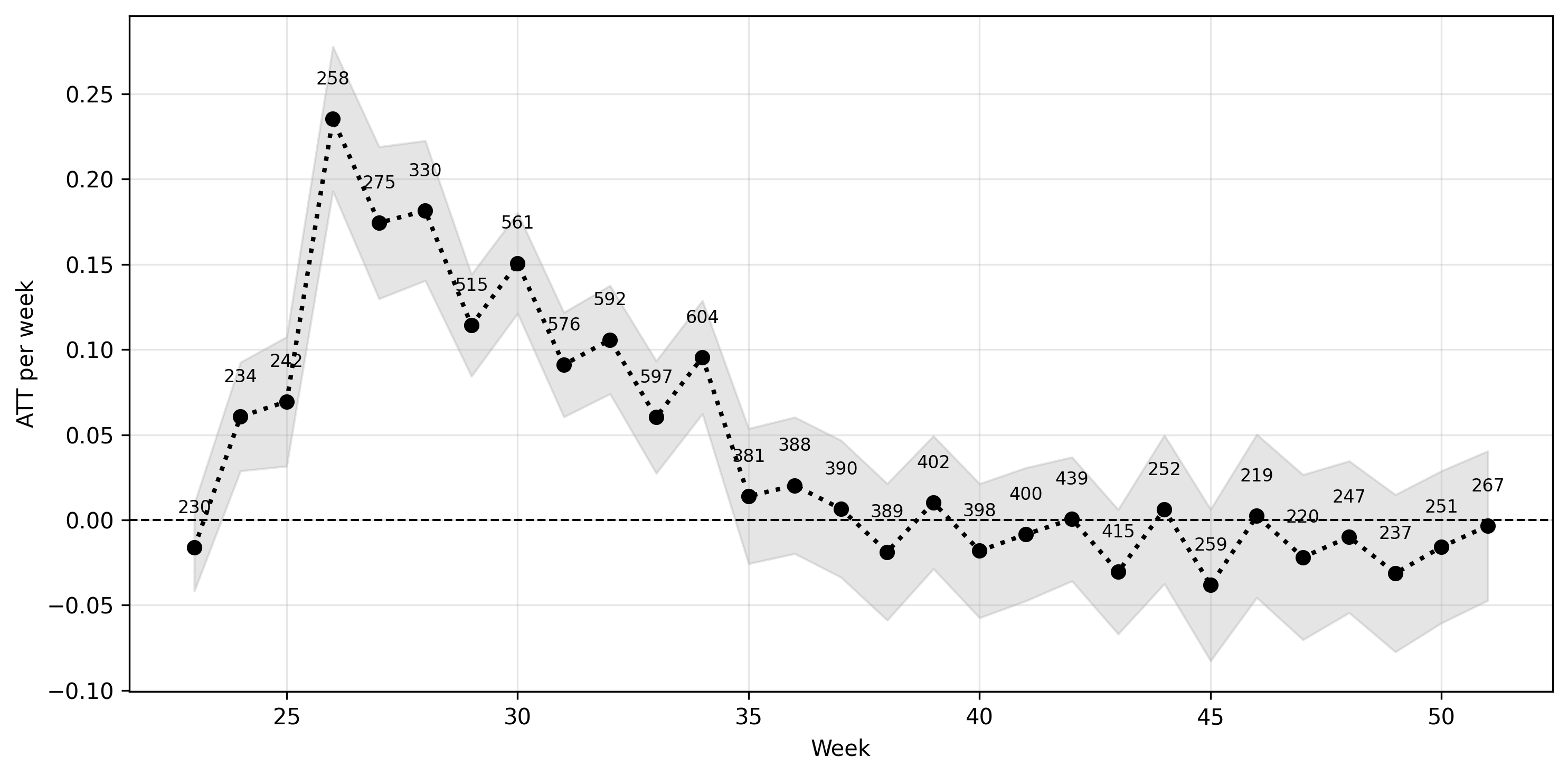}
  \caption{Time-specific ATT for $y$ with 95\% credible intervals, under GP-RBF-RBF specification.}
  \label{fig:cumulative-att}
\end{figure}

Figure~\ref{fig:cumulative-att} plots the evolution of the ATT for the final application. There is a noticable difference between the estimated causal effects corresponding to post treatment weeks for earlier weeks, compared to the later ones.

Taken together, these results suggest that the positive ATT estimates are primarily driven by the small systematic bias of the RBF (time + covariates) model, rather than by the effect of the policy intervention, because the average and cumulative treatment-effect intervals are similar in the validation and final estimates. However, to confirm this hypothesis, the results must be adjusted for systematic bias. Moreover, the results depend on the validity of the identifying assumptions in Section~\ref{sec:assumptions}. In particular, potential spillovers across stations or violations of the no-interference assumption could bias counterfactual predictions.

\chapter{Discussion}
\label{ch:discussion}

The empirical analyses presented in this study illustrate both the advantages and the practical challenges of using Exchangeable Gaussian Processes for policy evaluation. An important finding from both applications is that the choice of the covariance kernel is a critical factor in model performance, as our validation procedures demonstrate that no single specification is superior. The optimal choice depends on the structure of the underlying data.

In the Proposition~99 application, the synthetic difference-in-differences (SDID) benchmark achieved the lowest root mean squared error (RMSE) in the pre-treatment period, confirming its status as a robust estimator for this dataset. However, the Exchangeable GP models yielded comparable predictive accuracy. Moreover, the predictive intervals of the GP specifications were well-calibrated. This accurate calibration is vital for policy evaluation, as underestimated uncertainty can lead to false positives. Consequently, when we apply the model to the post-intervention period, we find that the posterior probability mass is concentrated below zero, providing strong evidence that the policy caused a reduction in consumption, a result that aligns with the established literature.

The application to Greek petrol stations presents a more complex scenario characterized by substantial unit heterogeneity. Despite the noisy nature of weekly station-level data, the Exchangeable GP framework provided strong predictive accuracy. However, the validation procedure revealed a systematic negative bias in the counterfactual predictions across all kernel specifications. While the magnitude of this bias was small relative to the variance of the outcome, it complicates the causal interpretation of the final results. In the final estimation, the model suggests a positive effect of audits on reported sales; yet, because the estimated effect size is similar in magnitude to the bias detected during validation, the results are inconclusive.

\chapter{Conclusion}
\label{ch:conclusion}

This paper has developed and applied a framework for panel data causal inference based on Exchangeable Gaussian Processes. The primary contribution of this work lies in the flexibility and computational efficiency of the proposed model, as well as its applicability in staggered adoption settings with large panel data. The GP prior allows for the learning of complex, nonlinear trends without parametric specification of functional forms. By sharing a single latent mean process and assuming unit specific deviations, we significantly reduce the number of parameters to be estimated, compared to other multi-task semi-parametric specifications (e.g. Linear model of coregionalization (LMC) with full rank), leading to more stable inference.

A key innovation of this study is the adaptation of this framework to staggered adoption designs with large number of training observations $n$. Traditional multi-task GP methods scale poorly ($ \mathcal{O}(n^3) $), making the computational cost too high for datasets with thousands of units. We introduced a subsampling approach that breaks the large-scale problem down into independent estimation tasks for each treated unit, using a random subset of controls. This strategy drastically reduces the computational cost, making Bayesian non-parametric inference feasible for large panels.

While this approach yields promising results, it opens important directions for future research. Future work should compare the predictive accuracy of the exchangeable model trained on the full dataset versus the subsampled approximation to assess the loss in predictive accuracy associated with subsampling controls. Such an analysis would help establish optimal guidelines for the size of the control subset required to maintain estimator consistency while preserving computational efficiency. Ultimately, the Exchangeable GP represents a powerful addition to the panel data causal inference toolkit, offering a balance between the flexibility of machine learning and the interpretability required for rigorous policy evaluation.

\clearpage

\renewcommand\thechapter{\Alph{chapter}}
\setcounter{chapter}{0}

\renewcommand\thefigure{\thechapter.\arabic{figure}}
\renewcommand\thetable{\thechapter.\arabic{table}}
\setcounter{figure}{0}
\setcounter{table}{0}

\renewcommand{\chapter}[1]{%
  \par
  \vspace{2em}
  \refstepcounter{chapter}%
  \noindent\textbf{\Large Appendix~\thechapter\quad #1}%
  \par\medskip
}


\chapter{Supplementary tables and figures}
\label{app:supplement}

\section{Example posterior predictive runs for petrol stations}
\label{app:example-runs}

In this section, we report two representative posterior predictive runs for the staggered-adoption application (Section~\ref{sec:appl2}) for treated units under the RBF exchangeable GP specifications discussed in Section~\ref{ch:exchangeable-gp}. These figures depict the posterior predictive distribution of the counterfactuals associated with the ``fake treatment.'' Each plot displays the observed outcome trajectory of a treated unit alongside the posterior predictive mean and the corresponding 95\% predictive interval. The dotted line indicates the ``fake treatment'' time.

\begin{figure}[htbp]
  \centering
  \includegraphics[width=0.9\textwidth]{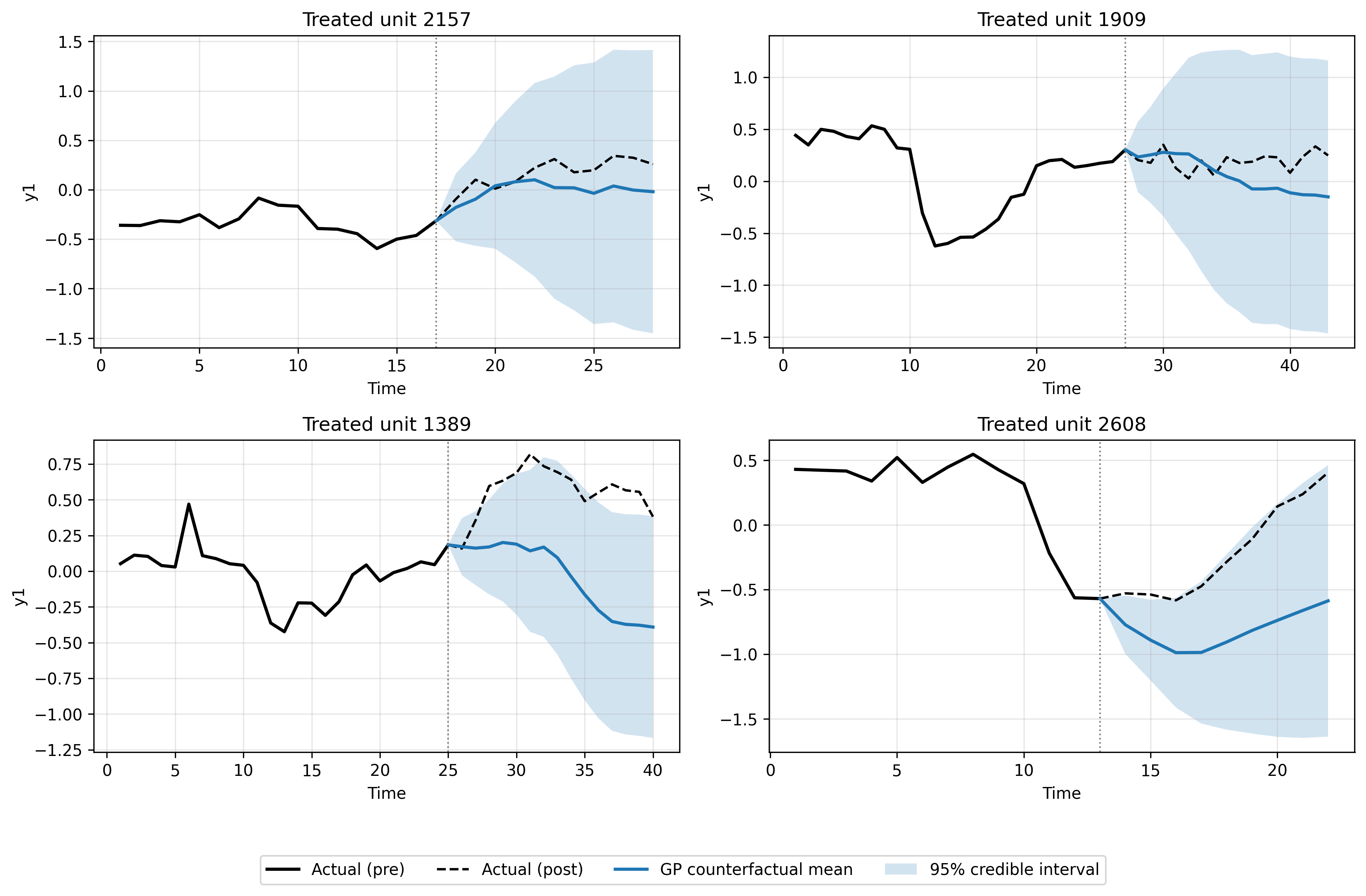}
  \caption{Example posterior predictive paths for four treated units under the RBF kernel specification on time for an exchangeable GP model.}
  \label{fig:example-run-rbf-time}
\end{figure}

\begin{figure}[htbp]
  \centering
  \includegraphics[width=0.9\textwidth]{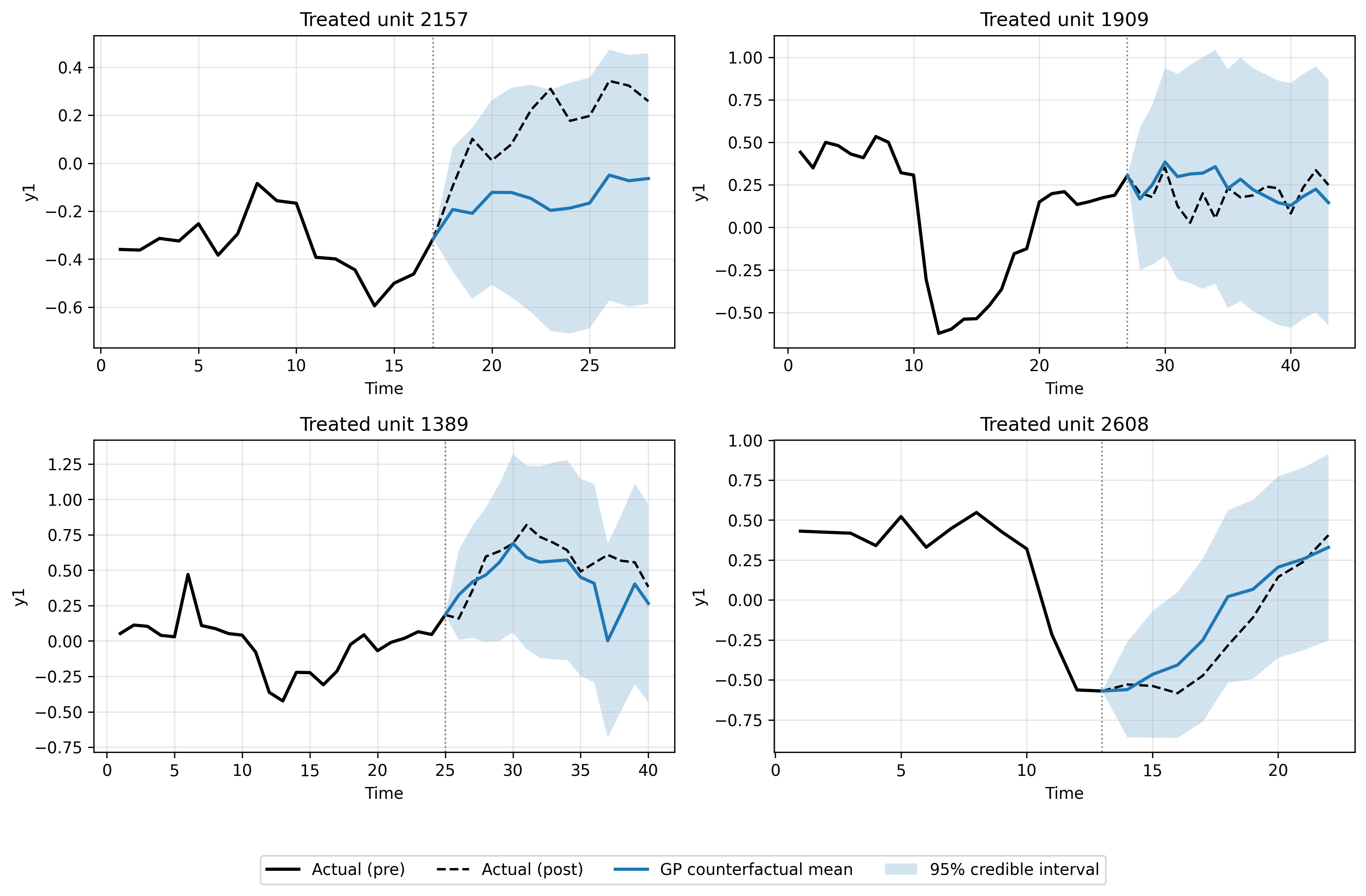}
  \caption{Example posterior predictive paths for four treated units under the RBF kernel specification on time and covariates for an exchangeable GP model.}
  \label{fig:example-run-rbf-time-covs}
\end{figure}
\FloatBarrier

\section{Horizon- and time-specific accuracy, and post-treatment ATT (RBF time + covariates) for Application 2}
\label{app:rbf-time-covs-results}

This section reports detailed diagnostics and treatment effect estimates for the RBF-time-covs specification in the staggered-adoption application (Section~\ref{sec:appl2}). We provide (i) horizon-specific prediction accuracy, (ii) time-specific prediction accuracy, and (iii) time-specific average treatment effects on the treated (ATT), each with corresponding uncertainty summaries.

In all of the tables, $n$ refers to the total number of predictions for that time or horizon. Table~\ref{tab:horizon-rbf-time-covs} reports prediction-accuracy metrics by forecast horizon (number of steps ahead) for outcome $y$ under the RBF time + covariates model. Coverage refers to the share of observed outcomes that fall within the 95\% posterior predictive interval. Table~\ref{tab:time-rbf-time-covs} reports week-by-week prediction accuracy for outcome $y$ under the same model. Table~\ref{tab:att-time-rbf-time-covs} reports week-specific ATT estimates for outcome $y$, along with 95\% credible intervals.

\begin{table}[p]
  \centering
  \caption{Horizon-specific prediction accuracy for $y$ (RBF time + covariates).}
  \label{tab:horizon-rbf-time-covs}
  \scriptsize
  \setlength{\tabcolsep}{4pt}
  \resizebox{0.5\textwidth}{!}{%
  \begin{tabular}{rcccc}
    \toprule
    Horizon & $n$ & RMSE & Bias & Coverage (95\%) \\
    \midrule
    1  & 864 & 0.161 & -0.002 & 96.41 \\
    2  & 864 & 0.231 & -0.005 & 95.25 \\
    3  & 864 & 0.269 & 0.018  & 94.68 \\
    4  & 864 & 0.290 & 0.003  & 95.95 \\
    5  & 864 & 0.335 & -0.020 & 95.60 \\
    6  & 864 & 0.351 & -0.030 & 96.06 \\
    7  & 864 & 0.386 & -0.062 & 95.02 \\
    8  & 864 & 0.371 & -0.059 & 96.06 \\
    9  & 864 & 0.398 & -0.067 & 94.68 \\
    10 & 630 & 0.375 & -0.032 & 96.51 \\
    11 & 589 & 0.375 & -0.036 & 96.10 \\
    12 & 303 & 0.370 & -0.050 & 96.37 \\
    13 & 267 & 0.310 & -0.049 & 95.88 \\
    14 & 246 & 0.357 & -0.051 & 95.93 \\
    15 & 220 & 0.307 & 0.012  & 96.36 \\
    16 & 150 & 0.354 & 0.018  & 96.00 \\
    17 & 90  & 0.395 & 0.017  & 95.56 \\
    18 & 41  & 0.212 & -0.015 & 100.00 \\
    \bottomrule
  \end{tabular}
  }%
\end{table}

\begin{table}[p]
  \centering
  \caption{Time-specific prediction accuracy for $y$ (RBF time + covariates).}
  \label{tab:time-rbf-time-covs}
  \tiny
  \setlength{\tabcolsep}{4pt}
  \resizebox{0.5\textwidth}{!}{%
  \begin{tabular}{rcccc}
    \toprule
    Week $t$ & $n$ & RMSE & Bias & Coverage (95\%) \\
    \midrule
    14 & 230 & 0.136 & -0.026 & 96.96 \\
    15 & 242 & 0.214 & -0.017 & 95.87 \\
    16 & 258 & 0.265 & 0.067  & 94.96 \\
    17 & 330 & 0.301 & 0.026  & 95.45 \\
    18 & 515 & 0.292 & -0.027 & 95.73 \\
    19 & 576 & 0.277 & -0.021 & 97.05 \\
    20 & 592 & 0.327 & -0.060 & 96.45 \\
    21 & 604 & 0.320 & -0.038 & 96.52 \\
    22 & 611 & 0.365 & -0.055 & 95.58 \\
    23 & 394 & 0.245 & -0.011 & 98.48 \\
    24 & 397 & 0.265 & -0.039 & 98.99 \\
    25 & 414 & 0.274 & -0.025 & 98.31 \\
    26 & 417 & 0.384 & -0.047 & 94.48 \\
    27 & 470 & 0.305 & 0.001  & 97.45 \\
    28 & 437 & 0.331 & -0.025 & 94.97 \\
    29 & 265 & 0.336 & 0.014  & 93.96 \\
    30 & 235 & 0.392 & -0.027 & 92.77 \\
    31 & 253 & 0.387 & -0.001 & 93.28 \\
    32 & 256 & 0.524 & -0.042 & 88.67 \\
    33 & 267 & 0.547 & -0.011 & 85.77 \\
    34 & 260 & 0.413 & -0.009 & 93.46 \\
    35 & 253 & 0.310 & -0.023 & 95.65 \\
    36 & 246 & 0.286 & -0.038 & 97.15 \\
    37 & 240 & 0.313 & -0.065 & 95.83 \\
    38 & 233 & 0.315 & -0.059 & 97.00 \\
    39 & 220 & 0.316 & -0.053 & 95.91 \\
    40 & 208 & 0.278 & -0.034 & 97.12 \\
    41 & 189 & 0.288 & -0.043 & 96.83 \\
    42 & 150 & 0.290 & -0.054 & 96.67 \\
    43 & 119 & 0.285 & -0.038 & 96.64 \\
    44 & 97  & 0.364 & -0.114 & 96.91 \\
    45 & 90  & 0.403 & 0.047  & 94.44 \\
    46 & 84  & 0.434 & 0.016  & 94.05 \\
    47 & 68  & 0.385 & 0.028  & 94.12 \\
    48 & 41  & 0.248 & -0.009 & 100.00 \\
    49 & 35  & 0.229 & 0.014  & 100.00 \\
    50 & 16  & 0.154 & -0.033 & 100.00 \\
    \bottomrule
  \end{tabular}
  }%
\end{table}

\begin{table}[htbp]
  \centering
  \caption{Time-specific ATT for $y$ (RBF time + covariates).}
  \label{tab:att-time-rbf-time-covs}
  \tiny
  \setlength{\tabcolsep}{4pt}
  \resizebox{0.7\textwidth}{!}{%
  \begin{tabular}{rccccc}
    \toprule
    Week $t$ & $n$ & Mean $\text{ATT}_t$ & SD of mean & 2.5\% CI & 97.5\% CI \\
    \midrule
    23 & 230 & -0.016 & 0.013 & -0.042 & 0.009 \\
    24 & 234 & 0.061 & 0.016 & 0.029 & 0.093 \\
    25 & 242 & 0.069 & 0.019 & 0.032 & 0.107 \\
    26 & 258 & 0.235 & 0.022 & 0.193 & 0.278 \\
    27 & 275 & 0.174 & 0.023 & 0.130 & 0.219 \\
    28 & 330 & 0.181 & 0.021 & 0.140 & 0.222 \\
    29 & 515 & 0.114 & 0.015 & 0.084 & 0.144 \\
    30 & 561 & 0.151 & 0.015 & 0.121 & 0.180 \\
    31 & 576 & 0.091 & 0.016 & 0.060 & 0.122 \\
    32 & 592 & 0.106 & 0.016 & 0.074 & 0.138 \\
    33 & 597 & 0.060 & 0.017 & 0.027 & 0.093 \\
    34 & 604 & 0.095 & 0.017 & 0.062 & 0.129 \\
    35 & 381 & 0.014 & 0.020 & -0.026 & 0.054 \\
    36 & 388 & 0.020 & 0.020 & -0.020 & 0.060 \\
    37 & 390 & 0.007 & 0.020 & -0.034 & 0.047 \\
    38 & 389 & -0.019 & 0.020 & -0.059 & 0.021 \\
    39 & 402 & 0.010 & 0.020 & -0.029 & 0.049 \\
    40 & 398 & -0.018 & 0.020 & -0.057 & 0.021 \\
    41 & 400 & -0.008 & 0.020 & -0.047 & 0.031 \\
    42 & 439 & 0.001 & 0.019 & -0.036 & 0.037 \\
    43 & 415 & -0.030 & 0.019 & -0.067 & 0.006 \\
    44 & 252 & 0.006 & 0.022 & -0.037 & 0.050 \\
    45 & 259 & -0.038 & 0.023 & -0.083 & 0.006 \\
    46 & 219 & 0.002 & 0.024 & -0.046 & 0.050 \\
    47 & 220 & -0.022 & 0.025 & -0.070 & 0.027 \\
    48 & 247 & -0.010 & 0.023 & -0.054 & 0.035 \\
    49 & 237 & -0.031 & 0.024 & -0.077 & 0.015 \\
    50 & 251 & -0.016 & 0.023 & -0.060 & 0.029 \\
    51 & 267 & -0.003 & 0.022 & -0.047 & 0.040 \\
    \bottomrule
  \end{tabular}
  }%
\end{table}
\label{app:horizon-accuracy}

\FloatBarrier

\clearpage
\bibliographystyle{elsarticle-harv}
\bibliography{bibliography}
\end{document}